\global\def\draftcontrol{0}

\def\versionno{ Taming Tadpoles }

\catcode`\@=11

\expandafter\ifx\csname draftcontrol\endcsname\relax\global\def\draftcontrol{0} 
\fi 

{\count255=\time\divide\count255 by 60 
	\xdef\hourmin{\number\count255} 
	\multiply\count255 by-60\advance\count255 by\time 
	\xdef\hourmin{\hourmin:\ifnum\count255<10 0\fi\the\count255}} 
\def\draftdate{\number\month/\number\day/\number\year\ \ \ \hourmin } 

\newcommand\makepapertitle{\par
	
	\begingroup 
	\renewcommand\thefootnote{\@fnsymbol\c@footnote}%
	\def\@makefnmark{\rlap{\@textsuperscript{\normalfont\@thefnmark}}}%
	\long\def\@makefntext##1{\parindent 1em\noindent 
		\hb@xt@1.8em{%
			\hss\@textsuperscript{\normalfont\@thefnmark}}##1}%
	\newpage 
	\global\@topnum\z@   
	\@makepapertitle 
	\thispagestyle{empty}\@thanks 
	\endgroup 
	\setcounter{footnote}{0}%
	\global\let\thanks\relax 
	\global\let\makepapertitle\relax 
	\global\let\@makepapertitle\relax 
	\global\let\@thanks\@empty 
	\global\let\@author\@empty 
	\global\let\@date\@empty 
	\global\let\@title\@empty 
	\global\let\title\relax 
	\global\let\author\relax 
	\global\let\date\relax 
	\global\let\and\relax 
	\def\version{\let\version\@version\@gobble} 
} 
\def\@makepapertitle{%
	\newpage 
	\ifnum\draftcontrol=1 {} 
	\version\versionno 
	\vskip 5.5em%
	\else 
	\hfill\hbox to 3.5cm {\parbox{4.5cm}{\@pubnum}\hss}%
	\vskip 6.5em%
	\fi 
	\begin{center}%
		\let \footnote \thanks 
		{\hskip -0\textwidth \hbox to 1\textwidth%
			{\centerline{\Large\bf{\noindent\@title}}}}%
		\vskip 2em%
		\hskip -0\textwidth\hbox to 1\textwidth%
		{\normalsize
			\lineskip .5em%
			\begin{tabular}[t]{c}%
				\@author 
			\end{tabular}\par}%
		\vskip 1.5em%
		{\@bstract}%
	\end{center}%
	\vfill
	\@date%
	\vskip 1.5em%
	\par 
} 

\gdef\@pubnum{} 
\def\pubnum#1{%
	\gdef\@pubnum{#1}} 

\gdef\@bstract{} 
\def\Abstract#1{%
	\gdef\@bstract{%
		\parbox{\textwidth-0pc}{%
			\centerline{\bf Abstract}\penalty1000 
			\noindent
			\renewcommand\baselinestretch{1.0} 
			{#1}}} 
} 

\gdef\@email{}
\def\email#1{%
	\gdef\@email{%
		Email: {\tt #1}}
}

\def\ps@paper{\let\@mkboth\@gobbletwo%
	\ifnum\draftcontrol=1 
	\def\@oddfoot{\hbox to \textwidth{\tiny \versionno \hfil\tiny\draftdate}%
		\hskip -\textwidth \hbox to \textwidth{\hfil\rm\thepage\hfil}}%
	\else\def\@oddfoot{\hbox to \textwidth{\hfil\rm\thepage\hfil}} 
	\fi 
	\let\@evenfoot\@oddfoot 
} 

\def\body{\clearpage 
	\pagestyle{paper} 
} 

\def\@version#1{\ifnum\draftcontrol=1 
	\typeout{}\typeout{#1}\typeout{} 
	\vskip3mm\centerline{\hbox{\fbox{\normalsize{\tt DRAFT -- #1 -- } 
				{\draftdate}}}}\vskip3mm 
	\fi} 
\let\version\@version 
\long\def\eqlabel#1{\ifnum\draftcontrol=1 
	\tag@false  
	\tag*{(\theequation) \hbox to -0.2cm{\hspace{0cm}\small{#1}\hss}} 
	\refstepcounter{equation}  
	\edef\@currentlabel{\theequation} 
	\ltx@label{#1}          
	\else 
	\label{#1} 
	\fi 
} 
\let\st@bibitem\@bibitem 
\let\st@lbibitem\@lbibitem 
\ifnum\draftcontrol=1 
\def\@bibitem#1{%
	\st@bibitem{#1}\a@@label{#1}\ignorespaces} 
\def\@lbibitem[#1]#2{%
	\st@lbibitem[#1]{#2}\a@@label{#2}\ignorespaces} 
\def\a@@label#1{%
	\gdef\a@lab{\smash{\normalfont\small#1}} 
	\ifvmode 
	\if@inlabel 
	\global\setbox\@labels\hbox{%
		\llap{\a@lab\let\a@lab\relax 
			\kern\@totalleftmargin\kern\marginparsep}%
		\box\@labels}%
	\fi 
	\fi} 
\fi 

\documentclass[12pt,letterpaper]{article} 

\usepackage{amsmath,bm,amsfonts,amssymb,array,calc,amsthm,rotating}
\usepackage{epsfig,psfrag}
\usepackage{pstool}
\usepackage{graphicx}
\usepackage{color}
\usepackage[colorlinks=false]{hyperref}
\usepackage{caption}
\usepackage{subcaption}

\renewcommand\baselinestretch{1.25} 
\renewcommand\section{\@startsection {section}{1}{\z@}%
	{-3.5ex \@plus -1ex \@minus -.2ex}%
	{2.3ex \@plus.2ex}%
	{\normalfont\large\bfseries}} 
\renewcommand\subsection{\@startsection{subsection}{2}{\z@}%
	{-3.25ex\@plus -1ex \@minus -.2ex}%
	{1.5ex \@plus .2ex}%
	{\normalfont\normalsize\bfseries}} 
\renewcommand\subsubsection{\@startsection{subsubsection}{3}{\z@}%
	{-3.25ex\@plus -1ex \@minus -.2ex}%
	{1.5ex \@plus .2ex}%
	{\normalfont\normalsize\it}} 
\renewcommand\paragraph{\@startsection{paragraph}{4}{\z@}%
	{-3.25ex\@plus -1ex \@minus -.2ex}%
	{1.5ex \@plus .2ex}%
	{\normalfont\normalsize\bf}} 
\renewcommand\subparagraph{\@startsection{subparagraph}{5}{\z@}%
	{-1.25ex\@plus -1ex \@minus -.2ex}%
	{0ex \@plus .2ex}%
	{\normalfont\normalsize\it}}

\numberwithin{equation}{section}

\long\def\@makecaption#1#2{%
	\vskip\abovecaptionskip
	\sbox\@tempboxa{{\bf #1:} #2}%
	\ifdim \wd\@tempboxa >\hsize
	{\small\bf #1:} {\small #2}\par
	\else
	\global \@minipagefalse
	\hb@xt@\hsize{\hfil\box\@tempboxa\hfil}%
	\fi
	\vskip\belowcaptionskip}

\renewcommand*\l@section[2]{%
	\ifnum \c@tocdepth >\z@
	\addpenalty\@secpenalty
	\addvspace{.5em \@plus\p@}%
	\setlength\@tempdima{1.5em}%
	\begingroup
	\parindent \z@ \rightskip \@pnumwidth
	\parfillskip -\@pnumwidth
	\leavevmode \bfseries
	\advance\leftskip\@tempdima
	\hskip -\leftskip
	#1\nobreak\hfil \nobreak\hb@xt@\@pnumwidth{\hss #2}\par
	\endgroup
	\fi}
\renewcommand*\l@subsection{\addvspace{.0em \@plus\p@}\@dottedtocline{2}{1.5em}{2.3em}}
\renewcommand*\l@subsubsection{\addvspace{-.2em \@plus\p@}\@dottedtocline{3}{3.8em}{3.2em}}

\definecolor{refcol}{rgb}{0.2,0.2,0.8}
\definecolor{eqcol}{rgb}{.6,0,0}
\definecolor{purple}{cmyk}{0,1,0,0}

\gdef\@citecolor{refcol}
\gdef\@linkcolor{eqcol}
\def\colorlinkspurple{\gdef\@urlcolor{purple}}
\def\colorlinksblue{\gdef\@urlcolor{blue}}
\def\colorlinksred{\gdef\@urlcolor{red}}

\def\revise#1       {\raisebox{-0em}{\rule{3pt}{1em}}%
	\marginpar{\raisebox{.5em}{\vrule width3pt\ 
			\vrule width0pt height 0pt depth0.5em 
			\hbox to 0cm{\hspace{0cm}{%
					\parbox[t]{4em}{\raggedright\footnotesize{#1}}}\hss}}}}

\def\calh         {{\cal H}}

\def\calm         {{\cal M}} 
 
\def\calo         {{\cal O}}

\def\calw         {{\cal W}}

\def\C            {{\mathbb C}} 

\def\zet          {{\mathbb Z}}

\def\del          {\partial} 
 
\def\sqr#1#2{{\vcenter{\vbox{\hrule height.#2pt   
				\hbox{\vrule width.#2pt height#1pt \kern#1pt 
					\vrule width.#2pt}\hrule height.#2pt}}}}

\DeclareMathOperator{\codim}{codim}
\DeclareMathOperator{\rank}{rank}

\newcommand{\be}{\begin{equation}}
\newcommand{\ee}{\end{equation}}
\newcommand{\ba}{\begin{eqnarray}}
\newcommand{\ea}{\end{eqnarray}}

\newcommand{\lp}{\left(}
\newcommand{\rp}{\right)}
\newcommand{\ls}{\left[}
\newcommand{\rs}{\right]}
\newcommand{\w}{\wedge}
\newcommand{\N}{\mathcal{N}}

\newcommand{\ZZ}{\mathbb{Z}}
\def\rmi{{\rm i}}

\def\jb{{\bar \jmath}}
\newcommand\bl{{\bf l}}
\newcommand\bx{{\bf x}}
\newcommand\bn{{\bf n}}

\allowdisplaybreaks

\catcode`\@=12 

\begin{document} 
	

\title{Fluxes, Vacua, and Tadpoles meet Landau-Ginzburg and Fermat}


\date{May 2025}

\author{
Katrin Becker$^{a}$, Eduardo Gonzalo$^{b}$,
Johannes Walcher$^{c}$, and
Timm Wrase$^{b}$
\ \\[0.4cm]
\hbox to \textwidth
{\centerline {\it $^{a}$ George P.\ and Cynthia Woods Mitchell Institute for Fundamental Physics and Astronomy}} \\
\it Texas A\&M University, College Station, TX 77843, U.S.A.
\\[0.2cm]
\it $^b$ Department of Physics \\
\it Lehigh University, Bethlehem, PA 18018, U.S.A.
\\[0.2cm]
\it $^c$ Mathematical Institute and Institute for Theoretical Physics\\
\it Ruprecht-Karls-Universit\"at Heidelberg, 69221 Heidelberg, Germany}

\Abstract{%
Type IIB flux vacua based on Landau-Ginzburg models without K\"ahler deformations provide
fully-controlled insights into the non-geometric and strongly-coupled string landscape. We show here that supersymmetric flux configurations at the Fermat point of the $1^9$ model, which were found long-time ago to saturate the orientifold tadpole, leave a number of massless fields, which however are not all flat directions of the superpotential at higher order. More generally, the rank of the Hessian of the superpotential is violating the refined tadpole conjecture but is compatible with a suitably formulated tadpole conjecture for all
fluxes that we found. Moreover, we describe new infinite families of supersymmetric 4d
$\mathcal{N}=1$ Minkowski and AdS vacua and confront them with several other swampland conjectures.}

\makepapertitle

\body

\version\versionno

\vskip 1em

\tableofcontents

\section{Introduction}
Moduli stabilization by fluxes has been a cornerstone of realistic string models since the advent of
the string landscape. Early investigations of the GKP construction \cite{Giddings:2001yu} such as refs.\
\cite{Giryavets:2003vd, Denef:2004dm, Denef:2005mm, Collinucci:2008pf} appeared to confirm the
expectation that a generic flux will stabilize all complex structure moduli of Calabi-Yau manifolds
in either type IIB or F-theory compactifications. Based on more recent studies, however, it was
argued that in models with a large number of complex structure moduli it should not be possible to
stabilize all of them using fluxes \cite{Bena:2020xrh,Bena:2021wyr,Bena:2021qty}. The basic idea,
known as the tadpole conjecture, is that the flux contribution to the D3-brane tadpole scales
linearly with the number of stabilized moduli, with a proportionality constant that leads to an
effective bound in many popular situations.
This argument is part of the swampland program (reviewed for example in
\cite{Palti:2019pca,vanBeest:2021lhn}), whose goal it is to determine what the low-energy effective
field theories are that can arise from a full-fledged theory of quantum gravity like string theory.

The relation between the size of the tadpole and the
number of stabilized moduli can easily be tested (and hence falsified) in many examples. Moreover,
with a somewhat more precise definition of the ``number of stabilized moduli'', the conjecture can
be stated essentially in classical Hodge theory, and could thus conceivably be proven rigorously
independent of complicated or unknown perturbative or non-perturbative quantum corrections.
Recent work along this line has provided evidence for the conjecture in the large complex structure
limits \cite{Betzler:2019kon, Plauschinn:2021hkp, Tsagkaris:2022apo} and also in more general asymptotic limits \cite{Grana:2022dfw}. A scenario
including a putative counterexample was presented in \cite{Marchesano:2021gyv}. However, this
counterexample was more recently challenged in \cite{Lust:2021xds,Grimm:2021ckh}.

The main aim of this paper is to shed light on the competition between the stabilization of moduli
and the size of the D3-brane tadpole in the deep interior of the moduli space of type IIB flux
compactification (see \cite{Braun:2020jrx} for related work in F-theory).
Our investigation is based on a Landau-Ginzburg orbifold describing a non-geometric compactification
with $h^{1,1}=0$, that was first studied with this purpose some 16 years ago in
\cite{Becker:2006ks,Becker:2007dn}. It was shown there that while the model is intrinsically
non-geometric, the standard Hodge theoretic formulas for the flux superpotential and tadpole
continue to apply, based on the holomorphic nature of the supersymmetric locus and thanks to
powerful non-renormalization theorems for the superpotential. Moreover, while the lattice of
supersymmetric fluxes at the Fermat point has such a large rank that brute force numerical searches
for ``short'' flux vectors compatible with tadpole cancellation are prohibitively expensive, some
explicit fluxes were found that lead to supersymmetric Minkowski and AdS vacua that are under full
control despite an $\calo(1)$ string coupling. However, the exact content of the low-energy theory
and the full set of supersymmetric fluxes remained unexplored at the time.

In this work, we will show first of all that in the Minkowski vacua of the $1^9$ Landau-Ginzburg
model presented in \cite{Becker:2006ks} in fact only a small subset of the $63$ complex structure
moduli (that survive the orientifold projection) obtain a mass as a consequence of the
flux. Secondly, we will present a more complete list of supersymmetric fluxes saturating the tadpole
and leading to 4d Minkowski vacua, and show that all of them contain a substantial number of
massless fields. Thirdly, based on the evaluation of the cubic (and higher-order) terms in the
superpotential, we show that not all of these massless fields correspond to truly flat directions,
although we are not able to show that all moduli are actually stabilized. Based on these insights,
we present a mathematically precise (if perhaps somewhat simplified) formulation of the tadpole
conjecture that can be tested non-trivially over the entire moduli space.

We then turn to other aspects of the swampland program, in which context the compactifications of
\cite{Becker:2006ks,Becker:2007dn} were revisited in the recent works \cite{Ishiguro:2021csu,
Bardzell:2022jfh}, focusing only on the stabilization of the three bulk complex structure moduli
(that are mirror dual to the untwisted K\"ahler moduli in the mirror dual toroidal type IIA
compactification). An intriguing result of \cite{Bardzell:2022jfh} was the presence of an infinite
family of SUSY Minkowski vacua. In this infinite family a quantized flux, which is unconstrained by
the tadpole, goes to infinity.
Here, we discover similar infinite families of Minkowski vacua that include all 
complex structure moduli of the model. One might then have to accept that an infinite family of 4d Minkowski solutions is part of the string landscape. This may sound contradictory to the standard lore that the landscape is finite, that is, that there is a finite number of vacua (and corresponding EFTs) below a certain energy cutoff \cite{Acharya:2006zw, Vafa:2005ui}. We argue that our infinite families of Minkowski vacua are consistent with the finiteness conjecture since we expect that for each family there is a tower of states becoming light.

Additionally, we revisit AdS solutions in these settings. There we find likewise new infinite
families of AdS solutions. The existence of these solutions was known based on a study of simple
models that restrict to the bulk moduli \cite{Becker:2007dn, Ishiguro:2021csu,
Bardzell:2022jfh}. Those families are reminiscent of the DGKT \cite{DeWolfe:2005uu,Camara:2005dc}
SUSY AdS vacua which are included in the mirror of these construction. Here we show that such
solutions also exist when taking all complex structure moduli into account. We present explicitly
two examples that have peculiar features that are relevant to the swampland program.

\section{Moduli stabilization in non-geometric backgrounds}\label{sec:review}

Following \cite{Becker:2006ks}, in this paper we focus on orientifolds of the $1^9$ Landau-Ginzburg
(LG) model, with worldsheet superpotential
\begin{equation}\label{eq:W19}
\mathcal{W} = \sum_{i=1}^{9} \Phi_i^3\,.
\end{equation} 
We orbifold by the $\mathbb{Z}_3$ symmetry $\Phi_i \rightarrow \omega \Phi_i$ where
$\omega = e^{\frac{2 \pi \rmi}{3}}$. It can be shown that the model is the mirror dual of a rigid
Calabi-Yau manifold ($T^{6}/\mathbb{Z}_3 \times \mathbb{Z}_3$). A basis for the $(c,c)$ primary
chiral superfields of the untwisted sector is given by the monomials $\Phi_{i}\Phi_{j}\Phi_{k}$
$i\neq j\neq k \neq i$. There are 84 of them and they can be identified as complex structure
moduli. In the untwisted sector of a LG model there is no $(a,c)$ sector, but there could be Kähler
moduli in the twisted sector. However, in the case of a $\mathbb{Z}_3$ orbifold one finds no
non-trivial $(a,c)$ primary superfields, so there are no Kähler moduli. Intuitively, the orbifold is
fixing the volume in string frame. Notice that this breaks S-duality in our setup.

\subsection{Orientifolds and fluxes}\label{ssec:ofluxes}

The different consistent orientifold projections were studied in \cite{Becker:2006ks}. Here we will
focus only on the first orientifold considered in \cite{Becker:2006ks}, which combines the
worldsheet parity operator with the operator $g_1$:
\begin{equation}\label{eq:orientifold}
g_1: (\Phi_1,\Phi_2,\Phi_3,...,\Phi_9) \rightarrow -(\Phi_2,\Phi_1,\Phi_3,...,\Phi_9) \,.
\end{equation}
This reduces the number of complex structure moduli down to 63: 7 coming from
$\Phi_{1} \Phi_{2} \Phi_{i}$, $i=3,4,\dots 9$, $\binom{7}{2}=21$ coming from
$(\Phi_{1}+ \Phi_{2}) \Phi_{i} \Phi_{j}$ and $\binom{7}{3}=35$ coming from
$ \Phi_{i} \Phi_{j} \Phi_{k}$.

Using results from \cite{Hori_2008}, the authors of \cite{Becker:2006ks} calculated the
        Ramond-Ramond charge of the crosscap state in the orientifold \eqref{eq:orientifold}, and
        showed that it amounts to 12 units of the one elementary B-brane in the model, which can
        naturally be addressed as a ``D3-brane'', keeping in mind that this is really an abuse of
        language because the model is intrinsically non-geometric.

Similarly, the possible Ramond-Ramond and Neveu-Schwarz fluxes, $F_3$ and $H_3$, can be studied by
consistency and comparison with the A-branes in the Landau-Ginzburg theory, which are the analogues
of supersymmetric three-cycles in ordinary Calabi-Yau compactifications. This gives on the one hand
their precise quantization condition, see eq.\ \eqref{eq:Ggamma} below, and on the other hand their
contribution to the Ramond-Ramond tadpole in the class of the orientifold. Including a possible
contribution from mobile D3-branes, the tadpole cancellation condition takes the standard form
\begin{equation}
\label{eq:tadpole}
N_{\rm flux} = \int_M F_3 \w H_3 = \frac{1}{\tau -\bar{\tau}} \int_M G
\w \bar{G} = \frac{N_{O3}}{2} - N_{D3} = 12 - N_{D3}\,,
\end{equation}
where we have introduced the axio-dilaton $\tau = C_0 + \rmi e^{-\phi}$ and the $G$-flux $G=F_3 - \tau H_3$.

We emphasize that the familiarity of the expression \eqref{eq:tadpole}, and other statements
in this section, should not belie the fact that their derivation and validity require delicate
consistency arguments from both worldsheet and spacetime considerations.

\subsection{Non-renormalization theorems}\label{ssec:nonrenormalization}

In particular, as explained in \cite{Becker:2006ks}, the flux superpotential is still given by the
usual GVW \cite{Dasgupta:1999ss, Gukov:1999ya} formula
\begin{equation}
\label{eq:GVW}
W=\int_M (F_3- \tau H_3) \w \Omega\,,
\end{equation}
with the understanding that the integral just refers to the abstract pairing in Landau-Ginzburg
cohomology, and, crucially, receives no perturbative or non-perturbative corrections, despite the
fact that the volume is fixed at string size by the orbifold, and the dilaton, as we will see, might be
stabilized at strong coupling. This was argued via the non-renormalization of the BPS tension of a
D5-brane domain wall but it also passes other non-trivial checks \cite{Becker:2006ks}.  One may
worry about brane instanton corrections. However, Euclidean D3-branes are absent since $h^{1,1}=0$
and D$(-1)$ instantons do not contribute in the large volume limit and are independent of the
internal volume. This absence of D$(-1)$ instantons seems also consistent with the recent paper by
Kim \cite{Kim:2022jvv} that finds no D$(-1)$ instantons if the O7-plane charges are locally
cancelled by D7-branes. Here, we have no O7-planes and D7-branes since $h^{1,1}=0$. One can also ask
why similar corrections should be absent in the type IIA mirror dual. For example, in the DGKT
construction \cite{DeWolfe:2005uu}, there is only one suitable 3-cycle, since $h^{2,1}=0$, and this
cycle is threaded by $H_3$-flux. So, one does not expect brane instanton corrections in the dual
setup either \cite{Marino:1999af}.

\subsection{Conditions for supersymmetric vacua}\label{ssec:reviewSUSY}

To study $\N=1$, $D=4$ supersymmetric vacua using the 4d effective action, we require, in addition
to the superpotential \eqref{eq:GVW}, also some knowledge of the K\"ahler potential $K$, which is
expected to receive both perturbative and non-perturbative string loop corrections. As pointed out
in \cite{Becker:2007dn} mirror symmetry implies that in the weak coupling, large complex structure
limit, the K\"ahler potential for the dilaton and the complex structure moduli is given by:
\begin{equation} \label{eq:K4}
K=-{\mathbf 4} \ln[ \tau-\bar \tau] - \ln[\rmi \int_M \Omega \wedge \overline{\Omega}].
\end{equation}
Note that this factor of 4 in front of the dilaton kinetic term does appear in the dimensional
reduction of the 10d type IIB supergravity action. However, for geometric compactifications the
proper 4d $\mathcal{N}=1$ chiral multiplets are related to the volume in Einstein frame
\cite{Grimm:2004uq}. The above orbifold we are doing can be thought of as fixing the volume in
string frame. Since ${\rm vol}_{6,{\rm string}} = e^{\frac32 \phi} {\rm vol}_{6,{\rm Einstein}}$ in
geometric compactification the 4 becomes a 1, the rest being absorbed into the term
$-2\ln( {\rm vol}_{6,{\rm Einstein}})$ in $K$. Given that our model is non-geometric this does not
happen. One could thus think of this factor as a small volume correction. However, more precisely
one should derive this factor of 4 using mirror symmetry \cite{Becker:2007dn}. This factor does
appear in the dimensional reduction of the 10d type IIA supergravity action on orientifolds of
Calabi-Yau manifolds \cite{Grimm:2004ua}. If we appropriately restrict our $H_3$-flux then our setup
is mirror dual to a type IIA string theory compactification on a rigid Calabi-Yau manifold. Thus, our
$K$ and $W$ as well as our solutions should agree with the type IIA results, which is the case if we
include this factor of 4. Note, that our setup is more general than the IIA compactifications since
upon mirror symmetry some of the $H_3$-flux quanta might become geometric or non-geometric fluxes on
the type IIA side.

In this work we study both SUSY Minkowski and AdS vacua. The factor of 4 is irrelevant for the
discussion of Minkowski solutions but it has important consequences for AdS solutions. In fact,
without it we would not find fluxes that are unconstrained by the tadpole cancellation condition and
that are expected from the type IIA mirror dual setup \cite{DeWolfe:2005uu,Camara:2005dc}. The
reason is that because of this factor of 4, the covariant derivative with respect to the dilaton has
an extra term:
\begin{equation}
D_{\tau}W=-\frac{1}{\tau - \overline{\tau}} \int_M \left( 3G + \overline{G} \right) \w \Omega =0\,.
\label{tau}
\end{equation}
Together with
\begin{equation}
D_{a}W= \int_M G \w \chi_{a} =0\,,
\label{chi}
\end{equation} 
where $\chi_{a}$ is a basis of $(2,1)$ harmonic forms, one finds that in a supersymmetric solution
$G$ can be written as:
\begin{equation}
G_{\text{SUSY}} = A^{a}\chi_{a} + A^{0} \left( -3 \Omega + \overline{\Omega} \right),
\label{eq:SUSYflux}
\end{equation} 
where $\Omega$ is the holomorphic 3-form and the $A$'s are complex coefficients. Notice that the
flux can have a $(3,0)$ component, which in turn implies that the supersymmetric flux is not
restricted to be imaginary self-dual (ISD). In fact, it provides an unbounded contribution to the
tadpole. The general formula for the tadpole is given by:
\begin{equation}
N_\text{flux}(G_{\text{SUSY}}) = \frac{81 \sqrt{3}}{2 \text{Im}\tau} \left( \sum_{a} |A^{a}|^2- 8|A^{0}|^2 \right)\,.
\label{flux}
\end{equation} 
Notice that the additional condition $W=0$ for Minkowski solutions implies that $G \in H^{2,1}(M)$, i.e., above we would have $A^0=0$ and $N_\text{flux}\geq 0$.

To describe and implement flux quantization, it is best to work with respect to an integral basis of
the middle cohomology lattice of the model, which measures charges of supersymmetric A-branes.  In
the Landau-Ginzburg model \eqref{eq:W19}, as reviewed in \cite{Becker:2006ks}, primitive cycles
$\Gamma_{\textbf{n}}$ are labelled by collections of $9$ integers $n_i\bmod 3$, $i=1,\ldots,9$,
which refer to the orientation of an elementary integration cycle for each variable $\Phi_i$ (see
appendix \ref{app:LGintegrals}). These $\Gamma_{\textbf{n}}$ are however not all linearly
independent and also subject to various identifications. In the orbifold, a basis is in
correspondence with the first $170$ non-negative integers written in binary notation with $9$
digits, $\textbf{n}=\left(n_1,n_2, \ldots, n_9\right)$, where $n_{i}=0,1$. The identification by the
orientifold \eqref{eq:orientifold} allows us to reduce the basis to $2^7=128$ elements labelled by
$\textbf{n}=\left(1,1, n_3, n_4, \ldots, n_9\right)$, where $n_{i}=0,1$. Let us denote their
Poincaré dual 3-forms as $\gamma_{\textbf{n}}$.  We can expand the flux 3-form in this basis:
\begin{equation}
G = \sum N^{\textbf{n}}\gamma_{\textbf{n}} - \tau \sum M^{\textbf{n}}\gamma_{\textbf{n}}\,,
\label{eq:Ggamma}
\end{equation} 
where $N^{\textbf{n}}$ and $M^{\textbf{n}}$ are integers, to ensure flux quantization.

Now our primary interest is finding fluxes such that we have a supersymmetric vacuum at the Fermat
point. That is, we need an appropriate choice of values for these integers $N_{\textbf{n}}$ and
$M_{\textbf{m}}$ such that equations \eqref{eq:SUSYflux} and \eqref{eq:Ggamma} are
compatible. To do this we remember that harmonic forms in LG models are represented by RR ground
states, which, in the model at hand, are again labelled by nine integers:\footnote{Note that we use
$\Omega_{\textbf{l}}$ to denote a basis of 3-forms and this should not be confused with the
holomorphic (3,0)-form $\Omega$ that does not have a subscript.}
\begin{equation}
\Omega_{\textbf{l}} \,\, \longleftrightarrow \,\,
\textbf{l}=\mid l^1 \dots, l^9 \rangle \quad {\rm with } \quad l^i = 1,2
\quad {\rm and } \quad \sum_{i=1}^9 l^i= 0  \quad {\rm mod} ~3\,.
\end{equation}
The classification of these states according to the four Hodge types of cohomology classes is shown
in table \ref{tablehodge}.
\begin{table}[t]
\begin{center}
\begin{tabular}{|c|c|c|c|c|}
\hline 
$\sum_{i}l_{i}$ & 9 & 12 & 15 & 18\tabularnewline
\hline 
$H^{(p,q)}$ & $H^{(3,0)}$ & $H^{(2,1)}$ & $H^{(1,2)}$ & $H^{(0,3)}$\tabularnewline
\hline 
\end{tabular}
\caption{Correspondence between harmonic 3-forms and RR ground states for our LG
model.} \label{tablehodge}
\end{center}
\end{table}
As we review in appendix \ref{app:LGintegrals}, their pairing with the integral classes
$\gamma_{\textbf{n}}$ is given by
\begin{equation}
\label{omegas}
\int \gamma_{\textbf{n}} \w \Omega_{\textbf{l}} = B_{\textbf{l}}
\;\omega^{{\textbf{n}}\cdot {\textbf{l}}} \qquad {\text{with} } \quad {\textbf{n}}\cdot {\textbf{l}}=\sum_i n_i l^i\,,
\end{equation}
where $\omega=e^{\frac{2\pi \rmi}{3}}$ and
$ B_{\textbf{l}} =\frac{1}{3^{9/2}} \prod_{i} \bigl( 1- \omega^{l^i} \bigr) \Gamma
(\frac{l^{i}}{3}) $ is a factor that drops out of equation \eqref{chi} which then implies that
\begin{equation}
\label{minkowski}
\sum_{\textbf{n}} \left( N^{\textbf{n}} - \tau M^{\textbf{n}} \right) \omega^{{\textbf{n}}\cdot {\textbf{l}}}=0
\qquad \text{for all $\textbf{l}$ with} \qquad \sum_i l^i =12 \,,
\end{equation}
while equation \eqref{tau} reduces to
\begin{equation}
\label{ads}
\sum_{\textbf{n}} \left( N^{\textbf{n}} - \tau M^{\textbf{n}} \right) \left(-3 \omega^{{\textbf{n}}\cdot
{\textbf{l}_{9}}}+\omega^{{\textbf{n}}\cdot {\textbf{l}_{18}}} \right)=0\,,
\end{equation}
where $\textbf{l}_{9}=\{1,1,1,1,1,1,1,1,1\}$ and $\textbf{l}_{18}=\{2,2,2,2,2,2,2,2,2\}$. These are
simple linear equations and we can solve them in full generality although we are dealing with a
large number of moduli.

\subsection{Higher-order derivatives of the superpotential}

Having found fluxes that are supersymmetric at a particular point in moduli space, the question
arises whether this actually leads to a stabilization of all the moduli, i.e., the absence of any
continuous zero-energy deformations of the vacuum. A sufficient condition for stability is that all
scalar fields corresponding to the erstwhile moduli be massive. However, even if in the presence of
massless fields, non-trivial interactions can stabilize the moduli at higher order in the
deformation parameters. In other words, the deformations could be marginal, but not exactly so.

In the language of {\it singularity theory}, a supersymmetric vacuum corresponds to a critical point
of the superpotential\footnote{In reality of course, the superpotential is not a function, but
merely a section of a particular line bundle over moduli space, as witnessed by the covariant
derivatives in \eqref{chi}.  Considerations of singularity theory being local are not directly
affected by this distinction after an appropriate choice of holomorphic coordinates.}
\eqref{eq:GVW}. The absence of continuous deformations means that the critical point is {\it
isolated}, while the absence of massless fields means that the critical point is {\it
non-degenerate}.  An example of a degenerate but isolated critical point is the origin $\Phi=0$ of
the superpotential $\calw=\Phi^3$.

In principle, the full dependence of the superpotential on all the moduli is encoded through
$\Omega$ in the formula \eqref{eq:GVW}. In practice, the explicit evaluation is generically
prohibitively complicated for more than a handful of variables. In our model at the Fermat point,
however, we can luckily evaluate all {\it derivatives} of the superpotential in terms of elementary
integrals, see appendix \ref{app:LGintegrals}, even if the full functional dependence remains
inaccessible.

We will focus on this momentarily, but wish to point out that the {\it second derivatives}
        of the superpotential (which in particular determine the masses of moduli) can in fact be
        evaluated much more easily for a generic Calabi-Yau through their relation with (what from
        its role in the heterotic string is known as) the {\it Yukawa coupling}. Namely, the Yukawa
        coupling captures the $(0,3)$-component of the third derivative of the holomorphic
        three-form
\begin{equation}
Y_{abc} = \int \Omega\wedge D_aD_bD_c\Omega \,,
\end{equation}
or equivalently the expansion of its second derivative in terms of the $(1,2)$-forms.  Referring
specifically to table 3 of \cite{Candelas:1990pi} the second derivatives of the flux superpotential
with respect to the complex structure moduli is given by
\begin{equation}
\label{eq:covcovW}
D_a D_b W = D_a \int G \w \chi_{b} = \int G \w \lp-\rmi e^{K} Y_{ab}^{\overline{c}}\chi_{\overline{c}} \rp\,.
\end{equation}
where we have also used that the flux, being dual to an integral cycle, is covariantly constant. To
reiterate, the interest of this observation is that the Yukawa coupling is on general grounds an
{\it algebraic function} on moduli space, and is therefore much more easily evaluated than the full
moduli dependence of $W$.

In the case at hand, labeling the complex structure moduli by the corresponding ${\bf l}$ vectors
whose entries sum to 12 according to table \ref{tablehodge}, and utilizing appendix
\ref{app:LGintegrals}, we find the following second derivatives of $W$
\begin{equation}
\label{eq:ddW}
D_{\bf l_a} D_{\bf l_b} W = \frac{1}{3^{9}}  \sum_{\textbf{n}} \left( N^{\textbf{n}}-\tau M^{\textbf{n}}
\right) \omega^{ \textbf{n}\cdot \left( \textbf{l}_{a}+\textbf{l}_{b}-\textbf{l}_{9} \right) }\prod_{i=1}^{9}
\bigl(1-\omega^{l_a^i+l_b^i-1}\bigr) \Gamma\left( \frac{l_a^i+l_b^i-1}{3} \right) \,.
\end{equation}
For the $\tau$ derivatives we find 
\ba
\label{eq:dtauda}
D_{\tau} D_{\bf l_a} W &=& -\frac{1}{\tau - \overline{\tau}} \int \left(3G + \overline{G} \right) \w \chi_{a}\,,\\
\label{eq:dtaudtau}
D_{\tau} D_{\tau} W &=& \frac{3}{\tau - \overline{\tau} } \int H_3 \w \Omega\,.
\ea
At the particular point $\tau=\omega$ we find
\ba
\label{eq:dtaudaexplicit}
D_{\tau} D_{\bf l_a} W \!\!&=&\!\! -\frac{1}{3^9(1+2 \omega)}  \sum_{\textbf{n}} \left(4
N^{\textbf{n}}-(-1+2\omega)M^{\textbf{n}} \right) \omega^{\textbf{n} \cdot \textbf{l}_a}
\prod_{i=1}^9 (1-\omega^{l_a^i}) \Gamma\left( \frac{l_a^i}{3} \right)\,,\qquad\\
\label{eq:dtaudtauexplicit}
D_{\tau} D_{\tau} W\!\! &=&\!\! \frac{1}{3^8(1+2 \omega)} \sum_{\textbf{n}} M^{\textbf{n}} \omega^{\textbf{n} \cdot \textbf{l}_9}
\prod_{i=1}^9 (1-\omega^{l_9^i}) \Gamma\left( \frac{l_9^i}{3} \right)\,.
\ea
When dealing with Minkowski vacua $G\in H^{2,1}(M)$ so the first equation simplifies to:
\begin{equation}
D_{\tau} D_{\bf l_a} W = -\frac{1}{3^9} \sum_{\textbf{n}}
 M^{\textbf{n}}\omega^{\textbf{n} \cdot \textbf{l}_a} \prod_{i=1}^9
(1-\omega^{l_a^i}) \Gamma\left( \frac{l_a^i}{3} \right) \,,\\
\end{equation}
and the last equation reduces to $D_{\tau} D_{\tau} W =0$. The multi-derivative of order $r$ with
respect to the moduli fields labelled by $\bl_1,\ldots,\bl_r$ is given by (see appendix \ref{app:LGintegrals} for
details):
\begin{equation}
\label{eq:orderr}
\begin{split}
\partial_{\bl_1} \partial_{\bl_2} \dots \partial_{\bl_r} W
=& \frac{1}{3^{9}} \sum_{\textbf{n}} \left( N^{\textbf{n}}-\tau M^{\textbf{n}} \right)
\omega^{ \textbf{n}\cdot \left( \sum_{\alpha} \textbf{l}_{\alpha}-(r-1)\textbf{l}_{9} \right) } \\
&  \times \prod_{i=1}^{9} \left(1-\omega^{ \sum_{\alpha} l_{\alpha}^i-(r-1)}\right)
\cdot \Gamma\left( \frac{\sum_{\alpha}^{r}l_{\alpha}^i-(r-1)}{3} \right) \,.
\end{split}
\end{equation}

\section{Around the tadpole conjecture}

In its original formulation \cite{Bena:2020xrh}, the tadpole conjecture states that ``the fluxes
which stabilize a large number, $h^{2,1}$ or $h^{3,1}$, of complex structure moduli of a Calabi-Yau
threefold or fourfold in a type IIB or F-theory compactification, respectively, make a positive
contribution to the D3-brane tadpole that grows at least linearly with the number of moduli'', i.e.,
there is a constant $\alpha$ such that for a large number of moduli
\begin{equation}
\label{eq:conjecture}
\text{(Flux tadpole)} > \alpha \times \text{(number of moduli)}
\end{equation}
The conjecture was motivated by a number of failed attempts to find models in which a large number
of moduli can be stabilized {\it explicitly} by fluxes. Moreover, based on these examples, it was
also proposed that more specifically, $\alpha$ is at least $1/3$. Since the formulation of the
conjecture, there have been a number of further tests and refinements, but to our knowledge no
substantial deviation or modification of \eqref{eq:conjecture} with $\alpha=1/3$ has yet been
observed.
A typical difficulty in either proving or disproving the conjecture appear to be somewhat fuzzy
hidden assumptions on the portion of moduli space in which stabilization is to be sought
for. Specifically, investigations such as \cite{Plauschinn:2021hkp,  
Marchesano:2021gyv,Lust:2021xds,Grana:2022dfw, Tsagkaris:2022apo} focus
around the large complex structure in order to control both K\"ahler and superpotential, and dismiss
any potential violations at the boundary of that region.
The methods of the present paper, as well as its precursors \cite{Becker:2006ks,Bardzell:2022jfh},
avoid some of these difficulties and, although they apply in a rather different region of moduli
space, offer at least the same level of control. Without taking sides, our concrete results motivate
us to raise two points which we feel need to be taken into account for a proper handle on the
tadpole conjecture.

\subsection{Hodge theoretic formulation}\label{ssec:hodge}

First of all, the conjecture \eqref{eq:conjecture} is stated without a precise definition of
``stabilization of moduli''. As our results illustrate, models located at special points in moduli
space with a high degree of symmetry will typically be missing mass terms for certain fields,
depending on their transformation properties, but these moduli can still be stabilized at
higher order in the deformation parameter.

Second, the conjecture focuses on the stabilization (however defined) of ``all of a large number of
complex structure moduli'' of a Calabi-Yau manifold by fluxes, with a universal constant $\alpha$.
In fact, and certainly from the mathematical point of view, it seems just as interesting to first
investigate the interplay between the size of the flux tadpole and the stabilization of only a
subset of the moduli {\it in a fixed model}, and worry later about the global problem and the
universality of $\alpha$.

To make progress, we propose\footnote{A similar point of view appears to be taken in
\cite{Grana:2022dfw}, but again restricted to the (strict) large complex structure limit, as well as
in the earlier work \cite{Braun:2020jrx} in F-theory. We also thank Hossein Movasati for
illuminating discussions, especially about the possible role of the rank of the supersymmetric
flux/Hodge lattice.} to investigate a version of the tadpole conjecture that can be formulated
purely in Hodge theoretic terms, and that makes sense over the entire moduli space. We do not claim
to capture all subtleties of moduli stabilization, especially those related to perturbative and
non-perturbative quantum corrections, or to the stabilization of K\"ahler moduli. However, we feel that
the actual problem is ``in the same universality class'' of landscape problems in the sense of
Douglas-Denef \cite{Denef_2004,Denef_2007}. Moreover, our formulation has the advantage of being
mathematically precise. We will spell out the proposal only for type IIB compactifications on
Calabi-Yau threefolds, because this is the class of models for which we have concrete results. We
point out, however, that the (obvious) reformulation for F-theory on Calabi-Yau fourfolds is even
more closely related to classical problems in Hodge theory.

Namely, given a Calabi-Yau threefold $Y$ with complex structure moduli space\footnote{It is
understood that the full construction involves an orientifold projection onto the invariant part of
the moduli space, fluxes have to be invariant, etc.}  $\calm_Y$ of dimension
$\dim\calm_Y=h^{2,1}(Y)\gg 1$, we may ask, for each point $z\in\calm_Y$, for the existence of
(non-zero) integral cohomology classes $F_3,H_3\in H^3(Y,\zet)$ and a value of the axio-dilaton
$\tau\in \calh$ in the upper half-plane, such that
\begin{equation}
G=F_3-\tau H_3\in H^{2,1}(Y_z) \oplus H^{0,3}(Y_z)
\end{equation}
is a supersymmetric flux with respect to the complex structure corresponding to $z$. For any such $G$,
the {\it flux tadpole}
\begin{equation}\label{eq:fluxtadpole}
Q(G) = \int_Y F_3\wedge H_3 = \frac{1}{\tau-\bar\tau} \int_Y G\wedge \bar G > 0 
\end{equation}
is a positive integer (if $G$ is non-zero). We will call the sublattice
\begin{equation}
\Lambda^{\rm SUSY}_{(\tau,z)} \subset H^3(Y,\zet)\oplus \tau H^3(Y,\zet)
\end{equation}
of such fluxes, with
$\rank \Lambda^{\rm SUSY}_{(\tau,z)}=: {\rm rk}(\tau,z)$ the {\it supersymmetric flux lattice}, and
the subset
\begin{equation}
\calm_Y^{\rm SUSY} =\{ (\tau,z)\in \calh\times \calm_Y \mid {\rm rk}(\tau,z)>0\} \subset \calh\times \calm_Y
\end{equation}
for which $\Lambda^{\rm SUSY}_{(\tau,z)}$ is non-trivial the {\it supersymmetric locus}. The fact
that $Q$ is positive definite gives $\Lambda^{\rm SUSY}_{(\tau,z)}$ an Euclidean structure.

$\calm_Y^{\rm SUSY}$ is in general a complicated space, with many components of possibly different
dimensions, as well as other singularities. We understand the gist of the tadpole conjecture as
relating the {\it codimension} of $\calm_Y^{\rm SUSY}$, as a measure for the number of ``stabilized
moduli'', to the smallest flux tadpole that engenders it. The issue related to the first point
raised above is that because of the singularities of the supersymmetric locus, there are in general
different notions of dimension. Among these, the {\it Zariski dimension}, defined in terms of the
maximal ideal $\mathfrak{m}$ at $(\tau,z)$ as
$\dim^Z_{(\tau,z)}(\calm^{\rm SUSY}_Y)= \dim \bigl(\mathfrak{m} /\mathfrak{m}^2\bigr)$, measures
essentially the number of fields that are left {\it massless} by the flux. On the other extreme, the
{\it Krull dimension} $\dim^K_{(\tau,z)}(\calm^{\rm SUSY}_Y)$, defined by the longest chain of prime
ideals of the local ring, corresponds to the maximal number of truly marginal deformations. The fact
that these are no more than the number of massless fields is the classical inequality
$\dim^K_{(\tau,z)}(\calm^{\rm SUSY}_Y) \le \dim^Z_{(\tau,z)}(\calm^{\rm SUSY}_Y)$, with equality
when the supersymmetric locus is smooth. The intuition is that a large rank of the flux lattice
corresponds to the intersection of many components of $\calm^{\rm SUSY}_Y$, and therefore leads to a
large discrepancy between $\dim^K$ and $\dim^Z$.

Now our results in the $1^9$ LG model at the Fermat point (where $\Lambda^{\rm SUSY}$ has extremely
large rank) indicate that while it is indeed difficult to find supersymmetric fluxes that make all
moduli massive with a bounded tadpole, higher order terms in the superpotential will stabilize
additional fields, although we have not been able to determine the exact number of surviving
continuous deformations. This leads us to propose that the tadpole conjecture of \cite{Bena:2020xrh}
should be extended and tested over the entire moduli space as the mathematical statement that {\it
the Zariski co-dimension of the supersymmetric locus of Calabi-Yau threefolds is bounded linearly by
the length of the shortest non-zero lattice vector},
\begin{equation}
\label{eq:zariskitad}
\codim^Z_{(\tau,z)}(\calm^{\rm SUSY}) \le \beta \cdot
\min\{ Q(G) \mid G\in \Lambda^{\rm SUSY}_{(\tau,z)}, G\neq 0 \}
\end{equation}
At this stage, the constant $\beta$ might depend on $Y$, but would according to the ``refined
tadpole conjecture'' \cite{Bena:2020xrh}, be uniformly bounded as\footnote{The first version of this paper did not have the factor of 2 in $\beta$ and concluded that this model does not seem to violate the refined tadpole conjecture. This factor of 2 arises from us defining the flux tadpole in equation \eqref{eq:fluxtadpole} in the covering space following \cite{Becker:2006ks}. The original tadpole conjecture paper~\cite{Bena:2020xrh} has in equation (2.1) a factor of 1/2 in front of the flux contribution, effectively counting the flux contribution in the quotient space.\\ The tadpole conjecture was further studied in non-geometric LG models in \cite{Becker:2024ijy, Becker:2024ayh, Rajaguru:2024emw}, where this point is further clarified. We thank Daniel Junghans for bringing this to our attention.} $\beta = (2\alpha)^{-1}\le 3/2$. 

In the following subsection \ref{ssec:hessianrank}, we explain the relationship between the number
of massive fields, i.e., the Zariski co-dimension, and the rank of the Hessian of the
superpotential, which is given in our model by eqs.\ \eqref{eq:ddW}, \eqref{eq:dtaudaexplicit},
\eqref{eq:dtaudtauexplicit}, and which can more generally be written in terms of the Yukawa coupling
as \eqref{eq:covcovW}. This last fact makes us particularly hopeful that the tadpole conjecture in
the form \eqref{eq:zariskitad} is amenable to a mathematical proof (or counterexample). Then, in
subsection \eqref{ssec:higherorder}, we turn to the analysis of the higher-order terms, given in our
model by \eqref{eq:orderr}. As alluded to above, this is in general a complicated problem in
(high-dimensional!)  singularity theory, and at this stage we will only describe an algorithm for
taking into account the first non-trivial correction.

\subsection{The rank of the mass matrix for Minkowski solutions}\label{ssec:hessianrank}
For 4d $\N=1$ theories the Lagrangian is given by
\be \mathcal{L} = -\frac12 K_{i\jb} \partial_\mu
\varphi^i \partial^\mu \bar{\varphi}^\jb - V\,, \quad \text{with} \quad V=e^K\lp K^{i\jb}D_i
W\overline{D_jW} -3|W|^2\rp\,.
\ee
Minkowski vacua satisfy the following relations
\be\label{eq:Minkowski}
W = \partial_\tau W = \partial_a W = 0\,, \quad \forall a=1,\ldots, h^{2,1}\,.
\ee
The Hessian matrix of the scalar potential for Minkowski vacua is then simply given by
$H_{i\jb} = \partial^i \partial_\jb V = e^K (\partial^i \partial_k W) K^{k \bar l} (\partial_{\bar
l} \partial_\jb W)$.\footnote{For Minkowski vacua one finds that the equations \eqref{eq:Minkowski}
imply that $\partial^i \partial_j V=0$.} To calculate the physical masses squared of the complex
scalar fields requires us to go to a canonical basis in field space using the diffeomorphism
${P^i}_j =\partial \varphi^i/\partial \hat{\varphi}^j$ defined such that
\be
-\frac12K_{i\jb} \partial_\mu \varphi^i \partial^\mu \bar{\varphi}^\jb =-\frac12K_{i\jb} {P_k}^i
\partial_\mu \hat{\varphi}^k {\bar{P}^{\jb}\vphantom{P}}_{\bar l} \partial^\mu
\bar{\hat{\varphi}}^{\bar l} = -\frac12 \delta_{i \jb}\, \partial_\mu \hat{\varphi}^i \partial^\mu
\bar{\hat{\varphi}}^\jb\,.
\ee
The masses squared are then given by the eigenvalues of the mass matrix
$M = (P^{-1})^T H \bar{P}^{-1}$. This can be rewritten as
\be
\label{eq:MassMatrxMink} M_{i \bar m}
=e^K [(P^{-1})^T (\partial\partial W)P^{-1}]_{ij} \delta^{j\bar{k}} [
(\bar{P}^{-1})^T(\bar{\partial}\bar{\partial} \bar{W})\bar{P}^{-1}]_{\bar{k} \bar{m}}\,.
\ee
We see that the masses squared are necessarily positive semidefinite. This ensures the stability of
the solution and is a result of the preserved supersymmetry. However, while instabilities in the
form of tachyons are forbidden, it is in principle possible that scalar fields are massless. To
answer the question of stability for a given solution with massless scalars would then require one
to calculate higher order terms in the scalar potential.\footnote{For example, a single real scalar
field with $V(\phi)=\phi^4$ would be massless but stabilized at $\phi=0$.} If some scalar fields in
4d $\N=1$ Minkowski vacua would remain flat directions, then one can only trust these vacua if one
can control \emph{all} corrections to the superpotential, because any kind of correction could lead
to a runaway potential for the flat directions. Given the non-renormalization theorems in
\cite{Becker:2006ks, Becker:2007dn} that we reviewed above in subsection
\ref{ssec:nonrenormalization}, we expect that these models do not receive any correction to the
superpotential (but only to the K\"ahler potential). So, the existence of these Minkowski vacua is
guaranteed independent of the corrections and higher order terms.

Let us return to the mass matrix above in equation \eqref{eq:MassMatrxMink}. It clearly involves in
addition to the superpotential also the inverse K\"ahler metric $K^{i\jb}$. Given that we have no
control over corrections to the K\"ahler potential we generically do not know what the masses of the
scalar fields are. However, we can ask more modest questions like whether all scalar fields are
massive or how many scalar fields are massless. This can actually be answered because the K\"ahler
metric appears in the kinetic terms and is therefore a positive definite matrix. For example, when
calculating the determinant of $M_{i \bar m}$ one finds that it is zero if and only if the
determinant of $\partial^i \partial_k W$ is zero \cite{Bardzell:2022jfh}. Thus, if the Hessian of
the superpotential has maximal rank then all scalar fields are massive. One can extend this argument
to show that the matrix rank of $M_{i \bar m}$ is the same as the matrix rank of
$\partial^i \partial_k W$: Assume there is a vector ${\bf a}$ in the nullspace of
$(\partial\partial W)$, i.e., $(\partial\partial W){\bf a}=0$. Then it follows from the definition
of $M_{i \bar m}$ in equation \eqref{eq:MassMatrxMink} that $\bar{P} \bar{{\bf a}}$ is in the
nullspace of $M$. Since $P$ is a diffeomorphism $\bar{P} \bar{{\bf a}}$ is non-zero whenever
${\bf a}$ is non-zero. Thus, $P$ provides a 1-1 map between nullvectors of $M$ and
$(\partial\partial W)$ and both matrices have therefore the same rank.

\subsection{Stabilization at higher order}\label{ssec:higherorder}

We have just seen that although we cannot calculate the physical masses of the moduli without
knowledge of the K\"ahler potential, the {\it number} of fields that remain massless in any given
Minkowski vacuum can be determined just based on the Hessian of the superpotential, i.e., the
quadratic terms in the expansion around the critical point \eqref{eq:Minkowski}. It is easy to see
that this extends to higher order as well: A continuous family of supersymmetric vacua just
corresponds to a flat direction of the superpotential, a problem which by analyticity can be studied
with knowledge of all derivatives at the critical point. In practice, this can be analyzed order by
order in the field expansion, which is still a very non-trivial problem, however.

To fix ideas, consider a theory with chiral fields $\varphi^i$, and superpotential $W$
expanded up to cubic terms around a critical point at the origin,
\begin{equation}
\label{eq:Wexp}
W = \frac 12 \sum_{i,j} H_{ij} \varphi^i\varphi^j + \frac 1{3!}\sum_{i,j,k} C_{ijk} \varphi^i\varphi^j\varphi^k
+ \text{{\it higher order terms}}\,.
\end{equation}
If the Hessian $H_{ij}$ does not have full rank, there are some massless fields, and we would like
to know how many of them correspond to true moduli, in particular, whether the critical point
equations
\begin{equation}
\label{eq:critexp}
\del_i W = H_{ij}\varphi^j + \frac 13 C_{ijk}\varphi^j\varphi^k + \cdots
\end{equation}
admit any continuous solutions.

In reality, this depends on the higher order terms.
For example, the superpotential $W=\frac 12 (\varphi-\psi^2)^2$ clearly has a flat direction along
$\varphi=\psi^2$.  However, in the expansion of $W$ up to cubic order
$\partial_\varphi W = \varphi-\psi^2$, $\partial_\psi W = -2\varphi\psi$ and if we treated these
truncated equations exactly, we would conclude that $\varphi=\psi=0$ is the only critical point. The
correct statement is that if we parametrize the deformation by the kernel of the quadratic term,
i.e., $\psi$, and eliminate $\varphi=\psi^2$ with the help of the first equation, the second
equation is $-2\psi^3=0$, and vanishes up to the order that we have kept track of, so we correctly
cannot conclude that the deformation is lifted.

In general, if we find that the equations \eqref{eq:critexp} do not vanish up to cubic order in the
independent fields once the massive fields have been eliminated, we can conclude that the
deformation space actually has smaller dimension than the kernel of $H_{ij}$. How many more fields
are stabilized at this order can be determined by solving a number of quadratic equations. To be
specific, assume that $H_{ij}$ has rank one, with $H_{11}=1$. The first equation $\del_1W=0$ is then
solved if
\begin{equation}
\label{eq:elimone}
\varphi^1 = - H_{1j} \varphi^j - \frac 13 C_{1jk}\varphi^j\varphi^k
\end{equation}
where the first sum is only over $j\neq 1$, but the second over all $j$, $k$. Since we are
only working up to third order in the independent fields ($\varphi^j$ for $j>1$), we can {\it
without harm replace} $\varphi^1$ with $-H_{1j}\varphi^j$ in the second term to solve $\del_1W$ up
to that order. Substituting this result in the remaining equations, the linear terms drop out
(because $H_{ij}$ had rank $1$) and we are left with a list of quadratic equations in only the
independent fields.  The reduction in dimension is a bit subtle, and not necessarily given by the
number of linearly independent quadrics, but as soon as one quadric is non-zero, we can conclude
that additional fields are stabilized in the full problem.

The generalization to Hessians of higher rank is straightforward, and we have implemented this for
the study of moduli stabilization in the $1^9$ Landau-Ginzburg model at the Fermat point. The
generalization to higher order in the fields if also fairly obvious, but we will leave it for
future work.

\section{Minkowski solutions at fixed coupling}\label{sec:Mink}

As we reviewed above, the superpotential does not receive perturbative or non-perturbative
corrections and thus supersymmetric Minkowski vacua are of particular interest in these 4d $\N=1$
theories. Originally Minkowski vacua in these non-geometric settings were studied in
\cite{Becker:2006ks} by including all complex structure moduli. Follow-up papers
\cite{Becker:2007dn, Ishiguro:2021csu, Bardzell:2022jfh} then restricted to the three bulk torus
complex structure moduli. It was shown in \cite{Becker:2007dn} that such vacua cannot arise at
parametrically large complex structure or weak coupling, thus confining them to a barely explored
part of the string landscape. In this section we will set the axio-dilaton to \be \tau = C_0 + \rmi
e^{-\phi} = \omega = e^{\frac{2\pi \rmi}{3}} = -\frac12 +\rmi \frac{\sqrt{3}}{2}\,.  \ee So, we have
a string coupling of order one and we will describe our attempts to systematically classify the
solutions of the $1^9$ model at this point in moduli space.

\subsection{Massless fields in old solutions}\label{ssec:OldSolutions}

The original work \cite{Becker:2006ks} presented three explicit Minkowski solutions of our model in
their section 4.5. These solutions were found essentially by happenstance, and no claim was
made as to their genericity or completeness. In view of the questions raised above, we have now
checked the rank of the corresponding Hessians for these three solutions. We find that the second
example of \cite{Becker:2006ks}, given in the $\Omega$-basis by\footnote{While we reproduced the $F_3$- and $H_3$-fluxes in \cite[Sec. 4.5]{Becker:2006ks}, we find a slightly different normalization and phase for the $G$-fluxes.}
\begin{align}\label{eq:solution8Om}
G_1 =\frac{1}{9} &\biggr[ -\Omega_{1,1,1,2,1,2,1,2,1}+\Omega_{1,1,1,2,1,2,1,1,2}+\Omega_{1,1,1,2,1,1,2,2,1} -\Omega_{1,1,1,2,1,1,2,1,2}\nonumber \\
&+\Omega_{1,1,1,1,2,2,1,2,1} -\Omega_{1,1,1,1,2,2,1,1,2}-\Omega_{1,1,1,1,2,1,2,2,1}+\Omega_{1,1,1,1,2,1,2,1,2} \biggr]
\end{align}
and which makes a contribution $N_{\rm flux}=8$ to the D3-brane tadpole, gives masses to 14 of the $64$ scalar
moduli.  The first example of \cite{Becker:2006ks},
\begin{align}\label{eq:solution4Om}
G_2 &=\frac{\rmi}{3\sqrt{3}} \biggr[ \Omega_{1,1,1,1,2,1,2,1,2}-\Omega_{1,1,1,1,2,1,2,2,1} -\Omega_{1,1,1,1,2,2,1,1,2}+\Omega_{1,1,1,1,2,2,1,2,1}\biggr]
\end{align}
with $N_{\rm flux}=12$ has 16 massive complex scalars. Finally, the third example:
\begin{align}\label{eq:solution12Om}
G_3=\frac{1}{9} & \biggr[ -\Omega_{1,1,1,2,2,2,1,1,1}-\Omega_{1,1,1,2,2,1,2,1,1}-\Omega_{1,1,1,2,2,1,1,2,1}+\Omega_{1,1,1,2,1,2,1,1,2}\\
&+\Omega_{1,1,1,2,1,1,2,1,2}+\Omega_{1,1,1,2,1,1,1,2,2}-\Omega_{1,1,1,1,2,2,2,1,1}-\Omega_{1,1,1,1,2,2,1,2,1}\\
&-\Omega_{1,1,1,1,2,1,2,2,1}+\Omega_{1,1,1,1,1,2,2,1,2}+\Omega_{1,1,1,1,1,2,1,2,2}+\Omega_{1,1,1,1,1,1,2,2,2} \biggr]
\end{align}
with $N_{\rm flux}=12$ has 22 massive complex scalars.

These numbers being rather small compared to the total number of moduli, we have set out to check on
the one hand whether any of the remaining massless fields are stabilized at higher order in the
field expansion, along the lines sketched in section \ref{ssec:higherorder}, and second to search
more systematically for Minkowski solutions in this model, hopefully covering all possibilities.

Along these lines, we have found that while in the vacua corresponding to $G_1$ and $G_3$ above, the
cubic terms in the superpotential do not lead to any further constraints on the massless fields, in
the vacuum corresponding to $G_2$, the $64-16=48$ massless scalars are subject to 10 linearly
independent quadratic equations at that order, so indeed some of them are actually stabilized and
not true moduli. These results in themselves suggest that this is not the full story, and that
quartic terms and beyond will lead to further stabilization. But we will leave this for future work
and instead turn to the systematic search.

\subsection{Symmetries and new solutions}\label{ssec:MinkSolutions}

The $1^9$ model specified in equation \eqref{eq:W19} enjoys an obvious $S^9$ permutation
symmetry. This symmetry is broken to $S^7$ by the orientifold action in equation
\eqref{eq:orientifold}. We thus need to study one representative of each $S^7$ orbit. However, this
is still a formidable task. For example, there are billions of ways of choosing 12
non-zero $A^a$ out of the 63 possible ones (cf. eqn. \eqref{eq:SUSYflux}) but the order of $S^7$ is only 5040.\footnote{This is
somewhat overestimating the actual problem since whenever we choose an $A^a$ that is not invariant
under the orientifold but rather gets mapped into $A^b$ then we have to turn on $A^b=A^a$.} Thus, we
somewhat randomly generate roughly eight hundred different Minkowski solutions that satisfy the tadpole
cancellation condition and have $N_{\rm flux}=12$. The maximum rank that we find for the mass matrix
is 26. This is consistent with the tadpole conjecture, which predicts that the number of massive
fields should be smaller than 36 given that the tadpole is 12. One example of this is given by:
\begin{align}\label{eq:rank26new}
G_4=-\frac{1}{9}& \biggr[ \Omega_{1,1,1,2,2,1,1,2,1}-\Omega_{1,1,2,1,2,1,1,2,1}-\Omega_{1,1,2,2,1,1,1,2,1} \nonumber+\Omega_{1,1,2,2,2,1,1,1,1} \\
& -\Omega_{1,2,1,2,2,1,1,1,1}-\Omega_{2,1,1,2,2,1,1,1,1}+\Omega_{1,2,2,1,1,1,1,2,1}+\Omega_{2,1,2,1,1,1,1,2,1}\nonumber \\
& -\Omega_{2,2,1,1,1,1,1,2,1} +\Omega_{2,2,1,1,2,1,1,1,1}+\Omega_{2,2,1,2,1,1,1,1,1}-\Omega_{2,2,2,1,1,1,1,1,1} \biggr]\,.
\end{align}
Next, we attempt to generalize the results to a full proof. We know that solutions which are related
by a permutation will have the same rank. Interestingly, we find that solutions with different
permutation symmetries can have the same rank. The solution $G_4$ has an
$ S_3\times \ZZ_2\times \ZZ_2 $ symmetry group. Another solution which also has rank 26 is
\begin{align}
G_5=\frac{\omega^2}{9}&\biggr[ \Omega_{1,1,2,2,2,1,1,1,1}-\Omega_{1,2,1,2,2,1,1,1,1}-\Omega_{2,1,1,2,2,1,1,1,1} \nonumber  +\Omega_{2,2,1,1,2,1,1,1,1}
\\
&+\Omega_{2,2,1,2,1,1,1,1,1}-\Omega_{2,2,2,1,1,1,1,1,1} +\Omega_{1,1,1,2,2,2,1,1,1}-\Omega_{1,1,2,1,2,2,1,1,1}
\nonumber \\
& - \Omega_{1,1,2,2,1,2,1,1,1}   +\Omega_{1,2,2,1,1,2,1,1,1}+ \Omega_{2,1,2,1,1,2,1,1,1}- \Omega_{2,2,1,1,1,2,1,1,1} \biggr]\,,
\end{align}
which has an $ S_3\times \ZZ_2 $ symmetry group. 

We also discovered new solutions that satisfy the tadpole condition with $N_{\text{flux}}=12$ and
have different rank for the mass matrix. For example, the following solution has 20 massive fields
respectively:
\begin{align}
G_7 =\frac{1}{9}&\biggr[-\Omega_{1,1,1,1,2,1,2,1,2}+\Omega_{1,1,1,1,2,1,2,2,1}+\Omega_{1,1,1,1,2,2,1,1,2}-\Omega_{1,1,1,1,2,2,1,2,1}	\nonumber \\
&+\omega(-\Omega_{1,2,1,1,1,1,2,1,2}+\Omega_{1,2,1,1,1,1,2,2,1}+\Omega_{1,2,1,1,1,2,1,1,2}-\Omega_{1,2,1,1,1,2,1,2,1}	\nonumber \\
&-\Omega_{2,1,1,1,1,1,2,1,2}+\Omega_{2,1,1,1,1,1,2,2,1}+\Omega_{2,1,1,1,1,2,1,1,2}-\Omega_{2,1,1,1,1,2,1,2,1})\biggr]\,,
\end{align}
and $G_8$ below has 18 massive fields
\begin{align}
G_8=\frac{1}{9}&\biggr[-\Omega_{1,1,1,1,2,1,2,1,2}+\Omega_{1,1,1,1,2,1,2,2,1}+\Omega_{1,1,1,1,2,2,1,1,2}-\Omega_{1,1,1,1,2,2,1,2,1}	\nonumber \\
&+\omega(-\Omega_{1,1,1,2,1,1,2,1,2}+\Omega_{1,1,1,2,1,1,2,2,1}+\Omega_{1,1,1,2,1,2,1,1,2}-\Omega_{1,1,1,2,1,2,1,2,1}	\nonumber \\
&-\Omega_{1,1,2,1,1,1,2,1,2}+\Omega_{1,1,2,1,1,1,2,2,1}+\Omega_{1,1,2,1,1,2,1,1,2}-\Omega_{1,1,2,1,1,2,1,2,1})\biggr]\,.
\end{align}
Summarizing, besides the rank 14 for the mass matrix that we only found for $N_{\rm flux}=8$ we
found via an extensive search of solutions with $N_{\rm flux}=12$ the ranks 16, 18, 20, 22 and
26. It is not clear to us whether there is a pattern emerging or not but it would be certainly
interesting to study this further.

We have also evaluated some higher-order corrections for these new solutions according to section
\ref{ssec:higherorder}. Including all terms in the superpotential that are cubic in the fields, we
find that while for $G_4$ and $G_5$, the number of massless fields is {\it not} reduced compared to
the second-order level, for $G_7$ we find $3$, and for $G_8$, $5$ additional quadrics in the
massless fields that are linearly independent over $\C$. This again is suggestive of an intriguing
pattern that we wish to study further in subsequent work.

Finally, we performed a search for flux configurations for which all scalar fields have a mass. This is of course the case for generic choices of fluxes, however, in those cases one overshoots the tadpole cancellation condition with the flux contribution by a lot. We tried to identify flux
configurations with a small contribution to the tadpole conjecture that stabilize all moduli but we did not manage to find solutions with $N_{\rm flux}$  smaller than 69. So, from this angle we did not find an inconsistency with the idea of the tadpole conjecture that in principle allows
full stabilization with $N_{\rm flux} > 64\cdot 2/3 \approx 42$.

\subsection{Constraint on the $G$-flux}\label{ssec:Gflux}
In order to tackle the question of what Minkowski solutions exist in our orbifold of the $1^9$ model, we need to understand potential constraints that reduce the parameters of a search. Here we summarize some constraints that we discovered when studying Minkowski vacua. 

We found that it is useful to write
\be\label{eq:GgenericAdS}
G_{\rm Mink} = \sum A^a \chi_a \in H^{2,1}(M)\,.
\ee
Following \cite[Section 4.3]{Becker:2006ks}, we can start by setting all but one of the $A^a$ equal to zero so that $G= A\, \Omega_{\bf l}$, where we switched to the $\Omega_{\bf l}$ basis discussed above in section \ref{ssec:reviewSUSY}. The quantization condition in equation \eqref{eq:Ggamma} then requires that for each $\textbf{n}$ in our basis there exist integers $N^{\bf n}$ and $M^{\bf n}$ such that
\be
\sqrt{3} A \omega^{\textbf{n}\cdot \textbf{l}} = N^{\bf n} -\tau M^{\bf n}\,.
\ee
Since we chose $\tau=\omega=e^{\frac{2\pi \rmi}{3}}$, we find that $|A|^2 \geq 1/3$. This then implies that (c.f. equation \eqref{flux} for $A^0=0$)
\be
N_{\rm flux} = 81 |A|^2 \geq 27\,.
\ee
This means, with only one non-zero $A^a$, it is impossible to find solutions within the tadpole bound of 12. One can repeat a similar argument for two non-zero $A^a$'s and finds that their contribution to the tadpole is at least 18 and therefore still too large. One can continue this analysis but it becomes more and more tedious to derive the analytic bounds. We are also more interested in an upper bound on the number of $A^a$ that we can turn on in order to make the analysis of this model tractable. 

Let us therefore turn on the generic $G$-flux in equation \eqref{eq:GgenericAdS} and write it explicitly as
\begin{equation}
G_{\rm Mink} = \sum_{\bf l} A^{\bf l} \Omega_{\bf l}= \sum_{\bf n} N^{\textbf{n}}\gamma_{\textbf{n}} - \tau \sum_{\bf n} M^{\textbf{n}}\gamma_{\textbf{n}}\,.
\end{equation} 
This allows us to write the 63 complex $A^a$ in terms of 126 real independent flux quanta. Denoting all the different integer combinations of flux quanta (that are themselves integer) schematically by $\mathbb{Z}$, we find via an explicit calculation that all $A^a$ can be brought into the form
\be
A^a = \frac19(\mathbb{Z} + \omega \mathbb{Z})\,.
\ee
The above constraint implies that each $A^a$ satisfies the constraint $|A^a|^2\geq 1/81$. This means that each non-zero $A^a$ we turn on has to contribute at least 1 to the tadpole
\be
N_{\rm flux} = 81 \sum_{a} |A^a|^2 \geq  \text{number of non-zero } A^a \,.
\ee
Thus, for our $1^9$ orientifold we can have at most 12 non-zero $A^a$ before violating the tadpole condition. Given that there exist Minkowski solutions with 12 non-zero $A^a$ and tadpole 12 (cf. eqn. \eqref{eq:solution12Om}), we know that one can saturate the bound in this case. Likewise, there is a solution with 8 non-zero $A^a$ and $N_{\rm flux}=8$ (cf. eqn. \eqref{eq:solution8Om}). So, again in this case one can saturate the bound. Given that in practice we only found solutions with $N_{\rm flux} \leq12$ for either four, eight or twelve non-zero $A^a$'s, we believe that there are further constraints that might potentially make a complete analysis of this model feasible. We leave this as an interesting challenge for the future.

\section{New infinite families}\label{sec:inffamilies}
The original paper \cite{Becker:2006ks} established the existence of Minkowski vacua in the full $1^9$ orientifold model, while follow-up papers restricted only to a subset of moduli. Interestingly, in one of these follow-up papers \cite{Bardzell:2022jfh} the existence of infinite families of Minkowski and AdS vacua for the torus bulk moduli was established. We will here now show that such infinite families also exist in the full model where we will include all 63 complex structure moduli and the axio-dilaton.

\subsection{Two infinite families of Minkowski vacua}

\subsubsection{Generalization of a previous solution}
We found it fairly straight forward to construct many different infinite families of Minkowski vacua, when we allow the axio-dilaton $\tau$ to vary. It was pointed out at the end of subsection 2.2 in \cite{Ishiguro:2021csu} that the effective action is not invariant under $SL(2,\mathbb{Z})$ symmetry. Nevertheless, the Minkowski vacuum equations in equation \eqref{eq:Minkowski} are transforming covariantly and infinite families arise from $SL(2,\mathbb{Z})$ transformations.\footnote{This is not the case for the AdS solutions we present below.} This means that $SL(2,\mathbb{Z})$ transformations seemingly generate new (infinite families of) solutions. Note that the argument for the breaking of the $SL(2,\mathbb{Z})$ symmetry in \cite{Ishiguro:2021csu} is due to the factor of 4 that appears in the K\"ahler potential in equation \eqref{eq:K4}. Given that the K\"ahler potential in equation \eqref{eq:K4} receives string loop corrections this argument seems strictly speaking only applicable at weak coupling and all our solutions are at strong coupling. So, while the scalar potential and for example the masses of the scalar fields in a weakly coupled Minkowski solution do change under $SL(2,\mathbb{Z})$ transformations, we cannot say for sure that the same is true for all the vacua in our infinite families, since we do not know the exact form of the corrections to the K\"ahler potential.

The family of solutions we present in this subsection is the first example given in \cite[Sec. 4.5]{Becker:2006ks} (see eqn. \eqref{eq:solution4Om} above) generalized by an arbitrary integer parameter $N \in \mathbb{Z}$. It's $G$-flux is given by
\ba
G &=& \frac{3N +\rmi \sqrt{3}(2-N)}{18 (N^2-N+1)} \ \bigl(\Omega_{1, 1, 1, 1, 2, 1, 2, 1, 2}-\Omega_{1, 1, 1, 1, 2, 1, 2, 2, 1} \cr
&&\qquad\qquad\qquad\qquad-\Omega_{1, 1, 1, 1, 2, 2, 1, 1, 2}+\Omega_{1, 1, 1, 1, 2, 2, 1, 2, 1}\bigr) \,.\qquad
\ea
The original solution in equation \eqref{eq:solution4Om} is recovered for $N=0$. The axio-dilaton for this infinite family is given by
\be\label{eq:tauexample1}
\tau =C_0 + \rmi e^{-\phi} = \frac{2 N-1}{2 \left(N^2-N+1\right)} + \rmi \frac{\sqrt{3}}{2 \left(N^2-N+1\right)}\,.
\ee
We see here, consistent with the argument in \cite{Becker:2007dn}, that we are always at strong coupling since
\be
e^{\phi} = \frac{2 \left(N^2-N+1\right)}{\sqrt{3}} \geq \frac{2}{\sqrt{3}}\,.
\ee
Let us recall from above that this is consistent with the fact that S-duality is broken in our setup by the orbifold that freezes the string frame volume. 

This infinite family of solutions has $N_{\rm flux}=12$ for any $N\in \mathbb{Z}$ so that the tadpole cancellation condition equation \eqref{eq:tadpole} is satisfied without requiring any D3-branes. We find that this particular solution has only 16 massive complex scalar fields. This is, maybe somewhat surprisingly, independent of the value of $N$. The explanation for this is that  the Hessian of the superpotential has zeros in most of its entries. The few non-zero entries are functions of $N$.

\subsubsection{A family with tadpole 12 and 26 massive scalar fields}
Let us present here another infinite family derived from the new solution we presented above in equation \eqref{eq:rank26new}. The family of solutions which has 26 massive complex scalar fields, again independent of the free parameter $N\in \mathbb{Z}$, has the $G$-flux
\ba
G &=& -\frac{(2-N) + \rmi \sqrt{3} N}{18 (N^2-N+1)}\, (\Omega_{1,1,1,2,2,1,1,2,1}-\Omega_{1,1,2,1,2,1,1,2,1}-\Omega_{1,1,2,2,1,1,1,2,1}\cr
&&\qquad\qquad\qquad\qquad\quad  +\Omega_{1,1,2,2,2,1,1,1,1} -\Omega_{1,2,1,2,2,1,1,1,1} - \Omega_{2,1,1,2,2,1,1,1,1}\cr 
&&\qquad\qquad\qquad\qquad\quad +\Omega_{1,2,2,1,1,1,1,2,1} +\Omega_{2,1,2,1,1,1,1,2,1} -\Omega_{2,2,1,1,1,1,1,2,1}\cr 
&&\qquad\qquad\qquad\qquad\quad +\Omega_{2,2,1,1,2,1,1,1,1} +\Omega_{2,2,1,2,1,1,1,1,1} -\Omega_{2,2,2,1,1,1,1,1,1}).\qquad
\ea
The original solution in equation \eqref{eq:rank26new} above is recovered for $N=0$. The axio-dilaton in this infinite family is
\be
\tau =C_0 + \rmi e^{-\phi} = -\frac{2N^2+1}{2 \left(N^2-N+1\right)} + \rmi \frac{\sqrt{3}}{2 \left(N^2-N+1\right)}\,.
\ee
Given that the expression for the dilaton is the same as in the previous infinite family, given in equation \eqref{eq:tauexample1} above, these solutions again only exist at strong coupling.

\subsection{Implications for the landscape and the swampland}

\subsubsection{The tadpole conjecture}
Given that we work with large number of $h^{2,1}=63$ complex structure moduli, the two infinite
families above, as well as the solutions discussed in subsections \ref{ssec:OldSolutions} and \ref{ssec:MinkSolutions}, serve as an interesting test of the tadpole conjecture. Let us stress again that this test is being performed in a strong coupling limit, away from the large complex structure point. In our solutions at the highly symmetric Fermat point we did find solutions with 26 massive complex structure moduli within the tadpole bound of 12. This leads to $12/26 \approx .46$, which is smaller than $2\alpha=2/3$ and thus provides a violation of the refined tadpole conjecture away from the boundary of moduli space. Likewise, the example in equation \eqref{eq:solution8Om} with $N_{\rm flux}=8$ and 14 massive fields leads to a violation because $8/14 \approx .57 < 2/3$. Given the large number of parameters in this model we have not been able to fully map out the solution space, so there is currently no proof preventing the existence of Minkowski vacua with more massive scalars. Let us also stress that we found that some scalars are stabilized by higher order terms. It would be very interesting to further study higher order stabilization. It is certainly possible and maybe even expected that higher order terms stabilize all moduli and thereby leave no flat directions. We hope to analyze this further in the future.

\subsubsection{The massless Minkowski conjecture}
Recently another swampland conjecture called the massless Minkowski conjecture was formulated in \cite{Andriot:2022yyj}. It states that 10d supergravity compactifications to 4d Minkowski space always have in the spectrum a massless scalar field that is a linear combination of the internal volume modulus, the dilaton and the volume moduli parameterizing the volume of the cycles transverse to the O-planes. This conjecture is connected to the tadpole conjecture above, which constrains the number of fields that can be stabilized. However, the massless Minkowski conjecture specifically only talks about massless fields and in principle allows a stabilization of all fields. It also does not require the presence of a large number of moduli and is thus clearly distinct from the tadpole conjecture in several ways.

While the conjecture stipulates a (geometric) compactification of 10d supergravity, which is not the case for the non-geometric model we study here, it is intriguing to note that we always find massless fields in all our Minkowski vacua. So, one could speculate that some version of this conjecture applies everywhere in moduli space.

As was already noted in \cite{Andriot:2022yyj}, our non-geometric models have no volume moduli and thus the massless scalars are among the complex structure moduli and the axion dilaton. We checked and find that generically the axio-dilaton is not a flat direction. Thus, the massless Minkowski conjecture would have to be generalized to allow any kind of modulus to be massless in a Minkowski vacuum to cover also our setting.

\subsubsection{Finiteness of vacua in quantum gravity}
Lastly, we would like to address here the apparent existence of an infinite number of 4d supersymmetric Minkowski vacua in our setup. There are arguments that quantum gravity should only allow for a finite number of vacua, see for example \cite{Acharya:2006zw}. This requirement was promoted to a swampland conjecture in \cite{Vafa:2005ui}. The finiteness of string theory vacua was recently generalized and put on more mathematical footing in the tameness conjecture \cite{Grimm:2021vpn}. This made it also possible to proof that there is only a finite number F-theory flux vacua with (imaginary) self-dual fluxes \cite{Bakker:2021uqw}. This might seem at first sight at odds with the existence of the infinite families of Minkowski vacua that we find above.

A precise statement about the finiteness of vacua was given in \cite[Section 4]{Hamada:2021yxy}. It says that below a fixed finite energy cutoff, there exist only a finite number of low energy effective field theories consistent with quantum gravity. For the counting one has to quotient by the moduli space. So, in our setup, if there would be flat directions we would have to quotient by them, however, each of the Minkowski vacua in the infinite families above would be a valid low energy effective theory below a \emph{certain} cutoff. The latter point is exactly the loophole that makes our infinite families consistent with the finiteness of vacua below a \emph{fixed} cutoff: In all infinite families the string coupling always runs to infinity. One therefore expects that an infinite tower of massive states becomes light in this limit. So, for any given \emph{fixed} cutoff we only have a finite number of vacua that are valid. It would be interesting to make this more precise. However, given that in our setup S-duality is broken, there are no weakly coupled dual solutions within our setup, making a more detailed study rather difficult.

\subsection{AdS vacua}\label{ssec:AdS}
It is also possible to study supersymmetric AdS solutions in this setting despite the fact that the K\"ahler potential receives unknown large corrections as discussed above in subsection \ref{ssec:nonrenormalization}. This was explained in \cite{Becker:2006ks} in two different ways: On the one hand we simply have to ensure that the $G$-flux is of the particular cohomology type given above in equation \eqref{eq:SUSYflux}. This choice is independent of $g_s$ corrections and therefore the existence of supersymmetric AdS solutions is unaffected by potential corrections. Another way of seeing this is by expanding the K\"ahler potential around the critical point and to allow for arbitrary corrections. One then finds that the corrections to $K$ are of the form $K \to K + \delta f(\varphi)+\overline{\delta f(\varphi)}$. This can be undone by a K\"ahler transformation $W \to e^{-\delta f(\varphi)} W$. Since the covariant derivative transform as $D_i W \to  e^{-\delta f(\varphi)}D_i W$ we see again that the existence of supersymmetric AdS solutions with $D_iW=0$ is not affected by arbitrary corrections.

In these non-geometric settings AdS vacua have been studied in \cite{Becker:2007dn, Ishiguro:2021csu, Bardzell:2022jfh}. It was shown in \cite{Becker:2007dn} that, restricting to the three torus bulk moduli, it is possible to find AdS vacua at parametrically large complex structure and parametrically weak coupling. Thus, for those solutions one has parametric control over all corrections and can trust the K\"ahler potential at large complex structure. It would be interesting to extend this study to include all 63 complex structure moduli and check that all can be in the large complex structure limit and are massive.\footnote{This is the expected result at least for AdS solutions that are mirror dual to the DGKT construction \cite{DeWolfe:2005uu}.} Here we do not explore this avenue but rather keep working with our Landau-Ginzburg model at the Fermat point. 

If the $G$-flux is chosen to be of the form in equation \eqref{eq:SUSYflux}:
\be
G_{\text{SUSY}} = A^{a}\chi_{a} + A^{0} \left( -3 \Omega + \overline{\Omega} \right)\,,
\ee
then we are automatically guaranteed to have a supersymmetric AdS solution if $A^0 \neq 0$. While we have shown in subsection \ref{ssec:Gflux} above that for Minkowski vacua we cannot have more than twelve non-zero $A^a$ without violating the tadpole this is not true for AdS solutions. In particular, a generic solution will have generically all $A^a$ non-zero and different. We have generated many such solutions with the constraint that they satisfy the tadpole cancellation in equation \eqref{eq:tadpole} with $N_{D3}=0$. This means the fluxes exactly cancel the contribution from the O3-planes. The $G$-flux for one such explicit solution with $\tau=e^{\frac{2\pi \rmi}{3}}$ is given explicitly in appendix \ref{app:GAdS}. This solution has $N_{\rm flux}=12$ and therefore satisfies the tadpole condition without any D3-branes. So, the only light fields are the 63 complex structure moduli and the axio-dilaton. 

The mass matrix for any 4d $\mathcal{N}=1$ supersymmetric AdS solution has off-diagonal entries that lead to a mass splitting between the two real scalars in the chiral multiplets. The Hessian of the scalar potential $V$ is given by
\ba
\partial_{i}\partial_{\jb} V &=&  e^K\ls (D_i D_k W) K^{k \bar l} (D_{\bar l} D_\jb \bar{W}) -2 K_{i\jb} |W|^2 \rs\,,\cr
\partial_{i}\partial_{j} V &=& - e^K (D_i D_j W) \bar{W}\,.
\ea
We see that the above involves the K\"ahler potential in a non-trivial way, which makes it difficult to say something definitive given that $K$ receives unknown correction at strong coupling. In the previous Minkowski solutions, we saw that the rank of the matrix $D_i D_j W$ was rather small and most entries in the matrix were zero in these examples. This told us that many scalar fields did not receive a mass through the fluxes that we have turned on. Here however we find that the $G$-flux example given in appendix \ref{app:GAdS} leads to rank 64 for the matrix $D_i D_j W$. This means that the scalar potential should involve all 64 complex scalars although the fluxes only contribute 12 to the tadpole. This is in stark contrast with the Minkowski solutions. Although we cannot calculate the masses explicitly here, given that an AdS solution exists and given that all of the scalar fields appear in the scalar potential, one might expect that generically all scalar fields will be massive in these solutions.\footnote{In AdS stability requires that the masses squared are all above the Breitenlohner-Freedman bound \cite{Breitenlohner:1982bm}. This is guaranteed for all our solutions because they are supersymmetric.}

While we have been able to generate many different such AdS solutions with $N_{\rm flux}=12$ and different mass matrix rank for $D_i D_j W$, we have not been easily able to extend these solutions to infinite families. We leave this as an interesting challenge for the future.

\subsubsection{A family with unbound tadpole}\label{ssec:AdS2}
Here we want to generalize an observation made in \cite{Bardzell:2022jfh} for the three bulk moduli to the full-fledged model at strong coupling: There are AdS solutions at large complex structure and weak coupling for which it is possible that $N_{\rm flux} \to - \infty$ \cite{Bardzell:2022jfh}. This would then require that $N_{D3}\to+\infty$ to satisfy the tadpole condition in equation \eqref{eq:tadpole}. For such solutions one then expects to have gauge groups with parametrically large rank. This is very different from Minkowski solutions where one expects a finite gauge group rank.

Here we present an infinite family of AdS solutions with two free parameters $N, M \in \mathbb{Z}$, $M\geq 0$. The $G$-flux is given by
\ba
G &=& \frac{(2M+1)}{3} \ls\bigl(\Omega_{1, 1, 1, 1, 2, 1, 2, 1, 2} + \Omega_{1, 1, 1, 2, 1, 2, 1, 2, 1} - \Omega_{1, 1, 2, 1, 1, 2, 2, 1, 1} +\Omega_{1, 1, 2, 2, 2, 1, 1, 1, 1} \right.\cr
&&\qquad \qquad \quad + \Omega_{1, 2, 1, 1, 1, 1, 1, 2, 2} + \Omega_{2, 1, 1, 1, 1, 1, 1, 2, 2} + \Omega_{1, 2, 1, 1, 1, 2, 2, 1, 1} + \Omega_{2, 1, 1, 1, 1, 2, 2, 1, 1} \cr 
&&\qquad \qquad \quad + \Omega_{1, 2, 1, 2, 2, 1, 1, 1, 1} + \Omega_{2, 1, 1, 2, 2, 1, 1, 1, 1} +\Omega_{2, 2, 2, 1, 1, 1, 1, 1, 1}\bigr)\cr 
&&\qquad \qquad \quad+ N (-3\ \Omega_{1, 1, 1, 1, 1, 1, 1, 1, 1} +\Omega_{2, 2, 2, 2, 2, 2, 2, 2,  2})\bigr]\,.\qquad
\ea
The axio-dilaton depends only on $M$ and for the string coupling to be positive we require that $M \geq 0$
\be
\tau = -\frac12 +\rmi \frac{\sqrt{3}\, (2 M + 1)}{2}\,. 
\ee
We see that the string coupling goes to zero for large positive $M$ values
\be
e^{\phi} = \frac{2}{\sqrt{3}\, (2 M + 1)}\,.
\ee
This means that in this solution we have parametric control over string loop corrections and we expect that all such corrections to the K\"ahler potential are suppressed in the large $M$ limit. The tadpole cancellation condition takes the form
\be
N_{\rm flux} + N_{D3} = -9 (2 M + 1) (72 N^2 -11) + N_{D3} = 12\,.
\ee
Given the constraint $M\geq0$ this can only be satisfied for $N \neq 0$. In that case, we see from the above that we need to add $N_{D3} = 12 +9 (2 M +1) (72 N^2-11)$ D3-branes. So, we expect solutions with an arbitrarily large gauge group rank. This is amusing but consistent with other examples in the literature \cite{Hanany:1997sa, Brunner:1997gf, Ahn:1998pb, Hanany:1997gh, Ferrara:1998vf}. It was argued in \cite{Lust:2019zwm} that AdS vacua do not allow for a scale separation between the AdS scale and an infinite tower of massive states. Thus, one cannot think of the gauge group as a genuine gauge group in AdS$_4$ but rather should think of it as a defect gauge group in a higher dimensional theory. The rank of such defect gauge theories is not bounded (as should be clear from a stack of Dp-branes in 10d flat space). In \cite{Bardzell:2022jfh} it was shown that similar solutions at weak coupling and large complex structure indeed seem to contain such a tower of light states. Actually, as was argued there, the open string moduli on the D3-branes would also lead to a species bound \cite{Veneziano:2001ah, Arkani-Hamed:2005zuc, Dvali:2007hz, Dvali:2007wp} that becomes small rather quickly.

\section{Conclusions}

Landau-Ginzburg techniques allow us to access string compactifications away from the large complex structure limit even for a very large number of moduli. In particular, at the Fermat point one has access to the values of the superpotential and all its derivatives. Furthermore, as was pointed out and used a long time ago in the papers \cite{Becker:2006ks, Becker:2007dn}, it is possible to find explicit Landau-Ginzburg models that are mirror dual to rigid Calabi-Yau and therefore have no K\"ahler moduli. Thus, compactifications of type IIB string theory on those models give rise to scalar potentials that can depend on all moduli and therefore in principle can give masses to all scalar fields. Here we have revisited those models to perform a more systematic study of supersymmetric Minkowski and supersymmetric AdS vacua.

One of the motivations for our study is the so-called tadpole conjecture \cite{Bena:2020xrh},
according to which, in violation of the well-informed intuition, it is in fact {\it not possible}
to stabilize all moduli using fluxes before these make an unacceptably large contribution to the
D3-brane tadpole. This question is usually studied in asymptotic limits near the boundary of moduli
space, where the effective action is best under control. Based on our results, which rely on
non-perturbative methods, we have proposed a version of the conjecture that is valid throughout
moduli space, and also makes a clear distinction between giving masses to moduli and stabilizing
them, potentially with a higher-order potential.

Among these results, we have for the first time determined how many complex structure moduli are massive in the previously known Minkowski solutions and we find that only 14, 16 or 22 out of 63 complex structure moduli and the axio-dilaton have a non-zero mass. Therefore, we carried out an extensive search for new Minkowski vacua and found many more solutions that have more massive scalar fields. The maximal value that we encountered is 26, so less than half of the scalar fields were massive. This violates the refined tadpole conjecture \cite{Bena:2020xrh}. However, our results agree with the general idea of the tadpole conjecture \cite{Bena:2020xrh} (in our reformulation) and thus provides a test of this conjecture away from the boundary of moduli space. While we find no violations of the tadpole conjecture in these Minkowski vacua, we have not been able to fully map out the moduli space. We also found that higher order terms in the scalar potential stabilize more scalar fields. Whether all fields can be stabilized or not is an important question that we plan to address in the future.

The recently proposed massless Minkowski conjecture \cite{Andriot:2022yyj} stipulates that supergravity compactifications to 4d Minkowski vacua always have massless scalars. While it does not strictly apply to our non-geometric setting, it is intriguing to note that we always find massless scalars in our Minkowski solutions. Thus we provide evidence for some version of the massless Minkowski conjecture away from the supergravity limit.

We have found that Minkowski vacua seem to generically come in infinite families. We presented
solutions in which a quantized flux, which does not appear in the tadpole cancellation condition,
can take arbitrary integer values. Increasing the absolute value of this flux quanta leads to
parametrically strong coupling. However, due to powerful renormalization theorems
\cite{Becker:2006ks, Becker:2007dn}, our solutions still exist at strong coupling. This seems at
odds with the believed finiteness of the number of vacua in string theory. However, we argue that
the strong coupling limit should signal that a tower of string states becomes light. This would mean
that the number of vacua below any given fixed cutoff scale is always finite, consistent with
previous expectations \cite{Acharya:2006zw, Vafa:2005ui}. The finiteness of string theory vacua has been recently revisited within the broader context of the tameness conjecture \cite{Grimm:2021vpn}. It was actually proven in \cite{Bakker:2021uqw} that there are only finitely many vacua in F-theory with (imaginary) self-dual fluxes. This should be in line with our families of solutions for which we argued that there is only a finite number below any fixed cut-off scale but it would be interesting to study this point further.

Lastly, we have also studied supersymmetric AdS solutions. Here our goal was two-fold: We have shown that for these AdS solutions it is possible to find superpotentials that depend on all 63 complex structure moduli and the axio-dilaton, while still only having a flux contribution to the tadpole that is equal to 12. We provide arguments that for those AdS vacua generically there seems to be no correlation between the number of massive scalar fields and the flux contribution to the tadpole cancellation condition. 

We have also presented one explicit infinite family of AdS solutions that goes to asymptotically weak coupling. In this limit the flux contribution goes to minus infinity and needs to be compensated for by D3-branes whose number goes to plus infinity. This thus leads to weakly coupled AdS solutions with gauge groups of arbitrarily large rank.

Given that several recent developments in the swampland program are guided by intuition from compactifications at large volume, large complex structure and weak coupling, it is of greatest importance to study string theory away from these limits. The Landau-Ginzburg models studied here, allow us to study non-geometric settings without any volume modulus, they allow us to work at small complex structure with many moduli and they even allow us to answer some questions at strong coupling. In this paper we have made several new interesting discoveries in this rather unexplored realm and many more are certain to await us.

\section*{Acknowledgements}
We would like to thank David Andriot, Thomas Grimm, Arthur Hebekcer, \'Alvaro Herr\'aez, Daniel Junghans, Manki Kim, Hossein Movasati, Eran Palti, Wati Taylor and Damian van de Heisteeg for helpful discussions. The work of K.B. is partially supported by the NSF grant PHY-2112859. The work of E.G. and T.W. is supported in part by the NSF grant
PHY-2013988. J.W.'s partial funding statement reads: This work is funded by the Deutsche Forschungsgemeinschaft (DFG, German Research Foundation) under Germany’s Excellence Strategy EXC 2181/1 — 390900948 (the Heidelberg STRUCTURES Excellence Cluster).

\appendix

\section{LG integrals}
\label{app:LGintegrals}

We summarize here the expansion of the Landau-Ginzburg periods
\begin{equation}
W_{\bn} = \int_{\Gamma_\bn} \Omega
\end{equation}
that enter the superpotential \eqref{eq:GVW}
\begin{equation}
W = \sum \bigl(N^\bn-\tau M^\bn\bigr) W_\bn
\end{equation}
when the $G$-flux is expanded according to \eqref{eq:Ggamma} in the basis dual to the $\Gamma_\bn$,
around the Fermat point.  This calculation is totally elementary and well-known to experts at least
from the days of \cite{Berglund_1994}.

\subsection{Single variable integrals}

\begin{figure}[t]
\begin{center}
\begin{subfigure}{.5\textwidth}
  \centering
  \includegraphics[width=.6\linewidth]{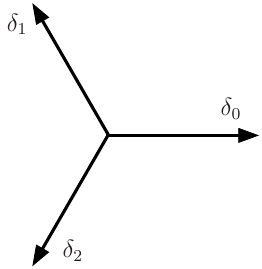}
\end{subfigure}%
\begin{subfigure}{.5\textwidth}
  \centering
  \includegraphics[width=.6\linewidth]{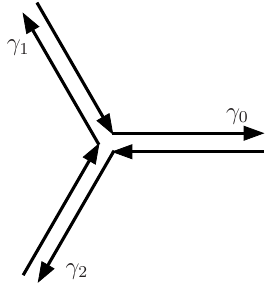}
\end{subfigure}
\caption{The integration cycles in a single variable Landau-Ginzburg model with worldsheet
superpotential $\calw=x^3$.}
\label{fig:piece}
\end{center}
\end{figure}

With reference to fig.\ \ref{fig:piece}, for $n=0,1,2\in\ZZ\bmod 3$ we let
$\delta_n :=\omega^n[0,\infty)\subset\C\ni x$ be the three independent rays along which $x^3$ tends
to real infinity, where $\omega=e^{2\pi \rmi/3}$. Then $\gamma_n:=\delta_n - \delta_{n+1}$ for
$n=0,1,2$ span the lattice of cycles, subject to the one relation
$\gamma_0+\gamma_1+\gamma_2=0$. For $l=1,2,\ldots$, we have
\begin{equation}
\int_{\delta_n} e^{-x^3}  x^{l-1} dx =
\omega^{n\cdot l} \cdot \frac{1}{3} \cdot \Gamma\bigl(\frac{l}{3}\bigr)
\end{equation}
and therefore
\begin{equation}
w_{n,l} := \int_{\gamma_n} e^{-x^3}  x^{l-1} dx = \omega^{n\cdot l} (1-\omega^l) \cdot \frac{1}{3}
\cdot \Gamma\bigl(\frac{l}{3}\bigr)\,.
\end{equation}

\subsection{Taylor coefficients}

Now, the deformation space of the $1^9$ LG model \eqref{eq:W19} is parametrized by local coordinates
$\{t^{\bl}, \bl = (l_1,\ldots, l_9)\in \{1,2\}^9, \sum l_i=12 \}$ via the worldsheet superpotential
\begin{equation}
\calw(t^\bl) = \sum_{i=1}^9 x_i^3 - \sum_{\bl} t^\bl \bx^{\bl-{\bf 1}}
\end{equation}
where $\bx^\bl = \prod_i (x_i)^{l_i} $ and ${\bf 1} = (1,1,1,1,1,1,1,1,1)$.
Then, for $\bn = (n_1,\ldots,n_9)\in \{0,1,2\}^9$, $\Gamma_\bn = \times_i \gamma_{n_i}$ we can write
the full moduli dependence of the period as
\begin{equation}
\begin{split}
W_\bn (t^\bl) & = \int_{\Gamma_\bn} e^{-\calw(t^\bl)} d^9x
= \int_{\Gamma_\bn} e^{-\calw(0)} \sum_{K=0}^\infty \frac{1}{K!} \bigl(
\textstyle\sum_{\bl} t^\bl \bx^{\bl-{\bf 1}} \bigr)^K d^9x
\\
&= \int_{\Gamma_\bn} e^{-\calw(0)} \sum_{k_\bl=0}^\infty
\prod_\bl \frac{\bigl(t^\bl\bigr)^{k_\bl}}{k_\bl!} \bigl(\bx^{\bl-{\bf 1}}\bigr)^{k_\bl} d^9x
\\
&= \sum_{k_\bl=0}^\infty \prod_\bl \frac{\bigl(t^\bl\bigr)^{k_\bl}}{k_\bl!}
\prod_{i} \int_{\gamma_{n_i}} e^{-x_i^3} \bigl(x_i\bigr)^{\sum_\bl k_\bl (l_i-1) } dx_i
\\
&= \sum_{k_\bl} \prod_\bl \frac{\bigl(t^\bl\bigr)^{k_\bl}}{k_\bl!} \prod_i w_{n_i, \sum_\bl k_\bl (l_i-1) + 1}
\end{split}
\end{equation}
Equivalently, and perhaps more simply, we can evaluate the $r$-th multi-derivative as
\begin{multline}
\eqlabel{taylorcoeffs}
\frac{\partial}{\partial t^{\bl_1}}
\frac{\partial}{\partial t^{\bl_2}}
\cdots
\frac{\partial}{\partial t^{\bl_r}}
W_\bn(0)
=
\int_{\Gamma_\bn} e^{-\calw(0)} \prod_{\alpha=1}^r \bx^{\bl_\alpha-{\bf 1}} d^9x
\\
=
\prod_i \int_{\gamma_{n_i}} e^{-x_i^3} \bigl(x_i\bigr)^{\sum_{\alpha=1}^r (l_\alpha^i-1) } dx_i
=
\prod_i w_{n_i,\sum_\alpha l_\alpha^i - r + 1}
\\
=
\frac{1}{3^9} \cdot
\omega^{\bn(\sum_\alpha\bl_\alpha-(r-1){\bf 1})}
\prod_i
\bigl(1-\omega^{\sum_\alpha l_\alpha^i -(r-1)}\bigr)
\cdot
\Gamma\Bigl(
\frac{\sum_\alpha l_\alpha^i - (r-1)}{3}\Bigr)
\end{multline}
which is the expression we used in the main text in equation \eqref{eq:orderr}.

\section{The $G$-flux for an AdS solution}\label{app:GAdS}
Here we give the explicit $G$-flux, discussed in subsection \ref{ssec:AdS} above. It describes an AdS solution that has $N_{\rm flux}=12$ and thus satisfies the tadpole cancellation condition without D3-branes. The value of the axio-dilaton is $\tau=e^{\frac{2\pi \rmi}{3}}$ and all 64 complex scalars appear in the Hessian of the superpotential since it has maximal rank 64. The $G$-flux is given by
\begin{align}
G &= \frac{1}{18} \biggr[ \left(15 -\rmi\sqrt{3}\right) \left( -3\Omega_{1,1,1,1,1,1,1,1,1} + \Omega_{2,2,2,2,2,2,2,2,2}  \right) \nonumber \\
&\rmi \left(3 \rmi+\sqrt{3}\right) \Omega_{1,1,1,1,1,1,2,2,2}  -4 \Omega_{1,2,1,1,2,1,1,1,2} +2 \Omega_{1,1,2,1,2,1,1,1,2} \nonumber \\
& +\rmi \left(3 \rmi+\sqrt{3}\right) \Omega_{1,1,1,1,1,2,1,2,2}  +\left(3+\rmi \sqrt{3}\right) \Omega_{1,1,1,1,1,2,2,1,2}+\rmi \left(3 \rmi+\sqrt{3}\right) \Omega_{1,1,1,1,1,2,2,2,1}  \nonumber \\
&+\left(-5-\rmi \sqrt{3}\right) \Omega_{1,1,1,1,2,1,1,2,2}+\left(3+\rmi \sqrt{3}\right) \Omega_{1,1,1,1,2,1,2,2,1}-2 \rmi \sqrt{3} \Omega_{1,1,1,1,2,2,1,1,2}  \nonumber \\
&+\left(6+2 \rmi \sqrt{3}\right) \Omega_{1,1,1,1,2,2,1,2,1}+2 \rmi \sqrt{3} \Omega_{1,1,1,2,1,1,1,2,2}+2 \Omega_{1,1,1,2,1,1,2,1,2} \nonumber \\
&+\rmi \left(3 \rmi+\sqrt{3}\right) \Omega_{1,1,1,2,1,1,2,2,1}+\rmi \left(5 \rmi+\sqrt{3}\right) \Omega_{1,1,1,2,1,2,1,2,1}+\left(1+\rmi \sqrt{3}\right) \Omega_{1,1,1,2,1,2,2,1,1}\nonumber \\
&+\left(2+2 \rmi \sqrt{3}\right) \Omega_{1,1,1,2,2,1,1,1,2} +4 \Omega_{1,1,1,2,2,1,1,2,1}+\left(-2-2 \rmi \sqrt{3}\right) \Omega_{1,1,1,2,2,1,2,1,1}\nonumber \\
&  +\left(-1-\rmi \sqrt{3}\right) \Omega_{1,1,1,2,2,2,1,1,1}+\rmi \left(3 \rmi+\sqrt{3}\right) \Omega_{1,1,2,1,1,1,1,2,2} 
+\left(3+\rmi \sqrt{3}\right) \Omega_{1,1,2,1,1,1,2,1,2}\nonumber \\
& +\rmi \left(3 \rmi+\sqrt{3}\right) \Omega_{1,1,2,1,1,1,2,2,1}+\left(3+\rmi \sqrt{3}\right) \Omega_{1,1,2,1,1,2,1,1,2}+\rmi \left(3 \rmi+\sqrt{3}\right) \Omega_{1,1,2,1,1,2,1,2,1}\nonumber \\
& +\left(3+\rmi \sqrt{3}\right) \Omega_{1,1,2,1,1,2,2,1,1}+2 \rmi \sqrt{3} \Omega_{1,1,2,1,2,1,1,2,1}+2 \Omega_{1,1,2,1,2,1,2,1,1}
\nonumber \\
& +\left(3+\rmi \sqrt{3}\right) \Omega_{1,1,2,1,2,2,1,1,1}+\left(1+\rmi \sqrt{3}\right) \Omega_{1,1,2,2,1,1,1,1,2} +\rmi \left(\rmi+\sqrt{3}\right) \Omega_{1,1,2,2,1,1,1,2,1}\nonumber \\
&+\left(1+\rmi \sqrt{3}\right) \Omega_{1,1,2,2,1,1,2,1,1}
+\left(1+\rmi \sqrt{3}\right) \Omega_{1,1,2,2,1,2,1,1,1}+\left(-5-\rmi \sqrt{3}\right) \Omega_{1,1,2,2,2,1,1,1,1} \nonumber \\
& -8 \Omega_{1,2,1,1,1,1,1,2,2}+\left(2+2 \rmi \sqrt{3}\right) \Omega_{1,2,1,1,1,1,2,1,2} +\rmi \left(3 \rmi+\sqrt{3}\right) \Omega_{1,2,1,1,1,1,2,2,1}\nonumber \\
& +2 \rmi \left(\rmi+\sqrt{3}\right) \Omega_{1,2,1,1,1,2,1,2,1}+\left(1+3 \rmi \sqrt{3}\right) \Omega_{1,2,1,1,1,2,2,1,1}+8 \Omega_{1,2,1,1,2,1,1,2,1}\nonumber \\
&+\left(-3-\rmi \sqrt{3}\right) \Omega_{1,2,1,1,2,1,2,1,1}+2 \Omega_{1,2,1,1,2,2,1,1,1}+\left(5+3 \rmi \sqrt{3}\right) \Omega_{1,2,1,2,1,1,1,1,2}\nonumber \\
&+\left(-1-\rmi \sqrt{3}\right) \Omega_{1,2,1,2,1,1,2,1,1}-4 \Omega_{1,2,1,2,1,2,1,1,1}+\left(4+4 \rmi \sqrt{3}\right) \Omega_{1,2,1,2,2,1,1,1,1}+\nonumber \\
&\left(3+3 \rmi \sqrt{3}\right) \Omega_{1,2,2,1,1,1,1,1,2}+2 \rmi \left(2 \rmi+\sqrt{3}\right) \Omega_{1,2,2,1,1,1,1,2,1}+\left(3+3 \rmi \sqrt{3}\right) \Omega_{1,2,2,1,1,1,2,1,1}\nonumber \\
&+\left(4+4 \rmi \sqrt{3}\right) \Omega_{1,2,2,1,1,2,1,1,1}+2 \rmi \left(2 \rmi+\sqrt{3}\right) \Omega_{1,2,2,1,2,1,1,1,1}+\rmi \left(3 \rmi+\sqrt{3}\right) \Omega_{1,2,2,2,1,1,1,1,1}\nonumber \\
&-8 \Omega_{2,1,1,1,1,1,1,2,2}+\left(2+2 \rmi \sqrt{3}\right) \Omega_{2,1,1,1,1,1,2,1,2}+\rmi \left(3 \rmi+\sqrt{3}\right) \Omega_{2,1,1,1,1,1,2,2,1}\nonumber \\
&+2 \rmi \left(\rmi+\sqrt{3}\right) \Omega_{2,1,1,1,1,2,1,2,1}+\left(1+3 \rmi \sqrt{3}\right) \Omega_{2,1,1,1,1,2,2,1,1}-4 \Omega_{2,1,1,1,2,1,1,1,2}\nonumber \\
&+\left(-3-\rmi \sqrt{3}\right) \Omega_{2,1,1,1,2,1,2,1,1}+2 \Omega_{2,1,1,1,2,2,1,1,1}+\left(5+3 \rmi \sqrt{3}\right) \Omega_{2,1,1,2,1,1,1,1,2}\nonumber \\
&+4 \Omega_{2,1,1,2,1,1,1,2,1}+\left(-1-\rmi \sqrt{3}\right) \Omega_{2,1,1,2,1,1,2,1,1}-4 \Omega_{2,1,1,2,1,2,1,1,1}\cr
&+\left(3+3 \rmi \sqrt{3}\right) \Omega_{2,1,2,1,1,1,1,1,2}+2 \rmi \left(2 \rmi+\sqrt{3}\right) \Omega_{2,1,2,1,1,1,1,2,1}+\left(3+3 \rmi \sqrt{3}\right) \Omega_{2,1,2,1,1,1,2,1,1}\cr
& +\left(4+4 \rmi \sqrt{3}\right) \Omega_{2,1,2,1,1,2,1,1,1}+2 \rmi \left(2 \rmi+\sqrt{3}\right) \Omega_{2,1,2,1,2,1,1,1,1}+\rmi \left(3 \rmi+\sqrt{3}\right) \Omega_{2,1,2,2,1,1,1,1,1}\nonumber \\ 
&+4 \Omega_{1,2,1,2,1,1,1,2,1}+8 \Omega_{2,1,1,1,2,1,1,2,1}+\left(4+4 \rmi \sqrt{3}\right) \Omega_{2,1,1,2,2,1,1,1,1}
\nonumber \\ & +\rmi \left(\rmi+\sqrt{3}\right) \Omega_{2,2,1,1,1,1,1,1,2}+8 \Omega_{2,2,1,1,1,1,1,2,1}-2 \Omega_{2,2,1,1,1,1,2,1,1}\nonumber \\
&+\rmi \left(\rmi+\sqrt{3} \right) \Omega_{2,2,1,1,1,2,1,1,1} +2 \rmi \sqrt{3} \Omega_{2,2,1,1,2,1,1,1,1}\nonumber \\
& +2 \left(5+2 \rmi \sqrt{3}\right) \Omega_{2,2,1,2,1,1,1,1,1}+\left(-2+4 \rmi \sqrt{3}\right) \Omega_{2,2,2,1,1,1,1,1,1}\biggr] \,.
\end{align}

\bibliographystyle{JHEP}
\bibliography{refs}

\providecommand{\href}[2]{#2}\begingroup\raggedright\begin{thebibliography}{10}

\bibitem{Giddings:2001yu}
S.B.~Giddings, S.~Kachru and J.~Polchinski, \emph{{Hierarchies from fluxes in
  string compactifications}},
  \href{https://doi.org/10.1103/PhysRevD.66.106006}{\emph{Phys. Rev. D}
  {\bfseries 66} (2002) 106006}
  [\href{https://arxiv.org/abs/hep-th/0105097}{{\ttfamily hep-th/0105097}}].

\bibitem{Giryavets:2003vd}
A.~Giryavets, S.~Kachru, P.K.~Tripathy and S.P.~Trivedi, \emph{{Flux
  compactifications on Calabi-Yau threefolds}},
  \href{https://doi.org/10.1088/1126-6708/2004/04/003}{\emph{JHEP} {\bfseries
  04} (2004) 003} [\href{https://arxiv.org/abs/hep-th/0312104}{{\ttfamily
  hep-th/0312104}}].

\bibitem{Denef:2004dm}
F.~Denef, M.R.~Douglas and B.~Florea, \emph{{Building a better racetrack}},
  \href{https://doi.org/10.1088/1126-6708/2004/06/034}{\emph{JHEP} {\bfseries
  06} (2004) 034} [\href{https://arxiv.org/abs/hep-th/0404257}{{\ttfamily
  hep-th/0404257}}].

\bibitem{Denef:2005mm}
F.~Denef, M.R.~Douglas, B.~Florea, A.~Grassi and S.~Kachru, \emph{{Fixing all
  moduli in a simple f-theory compactification}},
  \href{https://doi.org/10.4310/ATMP.2005.v9.n6.a1}{\emph{Adv. Theor. Math.
  Phys.} {\bfseries 9} (2005) 861}
  [\href{https://arxiv.org/abs/hep-th/0503124}{{\ttfamily hep-th/0503124}}].

\bibitem{Collinucci:2008pf}
A.~Collinucci, F.~Denef and M.~Esole, \emph{{D-brane Deconstructions in IIB
  Orientifolds}},
  \href{https://doi.org/10.1088/1126-6708/2009/02/005}{\emph{JHEP} {\bfseries
  02} (2009) 005} [\href{https://arxiv.org/abs/0805.1573}{{\ttfamily
  0805.1573}}].

\bibitem{Bena:2020xrh}
I.~Bena, J.~Bl\r{a}b\"ack, M.~Gra\~na and S.~L\"ust, \emph{{The tadpole
  problem}}, \href{https://doi.org/10.1007/JHEP11(2021)223}{\emph{JHEP}
  {\bfseries 11} (2021) 223}
  [\href{https://arxiv.org/abs/2010.10519}{{\ttfamily 2010.10519}}].

\bibitem{Bena:2021wyr}
I.~Bena, J.~Bl\r{a}b\"ack, M.~Gra\~na and S.~L\"ust, \emph{{Algorithmically
  Solving the Tadpole Problem}},
  \href{https://doi.org/10.1007/s00006-021-01189-6}{\emph{Adv. Appl. Clifford
  Algebras} {\bfseries 32} (2022) 7}
  [\href{https://arxiv.org/abs/2103.03250}{{\ttfamily 2103.03250}}].

\bibitem{Bena:2021qty}
I.~Bena, C.~Brodie and M.~Gra\~na, \emph{{D7 moduli stabilization: the tadpole
  menace}}, \href{https://doi.org/10.1007/JHEP01(2022)138}{\emph{JHEP}
  {\bfseries 01} (2022) 138}
  [\href{https://arxiv.org/abs/2112.00013}{{\ttfamily 2112.00013}}].

\bibitem{Palti:2019pca}
E.~Palti, \emph{{The Swampland: Introduction and Review}},
  \href{https://doi.org/10.1002/prop.201900037}{\emph{Fortsch. Phys.}
  {\bfseries 67} (2019) 1900037}
  [\href{https://arxiv.org/abs/1903.06239}{{\ttfamily 1903.06239}}].

\bibitem{vanBeest:2021lhn}
M.~van Beest, J.~Calder\'on-Infante, D.~Mirfendereski and I.~Valenzuela,
  \emph{{Lectures on the Swampland Program in String Compactifications}},
  \href{https://arxiv.org/abs/2102.01111}{{\ttfamily 2102.01111}}.

\bibitem{Betzler:2019kon}
P.~Betzler and E.~Plauschinn, \emph{{Type IIB flux vacua and tadpole
  cancellation}}, \href{https://doi.org/10.1002/prop.201900065}{\emph{Fortsch.
  Phys.} {\bfseries 67} (2019) 1900065}
  [\href{https://arxiv.org/abs/1905.08823}{{\ttfamily 1905.08823}}].

\bibitem{Plauschinn:2021hkp}
E.~Plauschinn, \emph{{The tadpole conjecture at large complex-structure}},
  \href{https://doi.org/10.1007/JHEP02(2022)206}{\emph{JHEP} {\bfseries 02}
  (2022) 206} [\href{https://arxiv.org/abs/2109.00029}{{\ttfamily
  2109.00029}}].

\bibitem{Tsagkaris:2022apo}
K.~Tsagkaris and E.~Plauschinn, \emph{{Moduli stabilization in type IIB
  orientifolds at $h^{2,1}=50$}},
  \href{https://arxiv.org/abs/2207.13721}{{\ttfamily 2207.13721}}.

\bibitem{Grana:2022dfw}
M.~Gra\~na, T.W.~Grimm, D.~van~de Heisteeg, A.~Herraez and E.~Plauschinn,
  \emph{{The tadpole conjecture in asymptotic limits}},
  \href{https://doi.org/10.1007/JHEP08(2022)237}{\emph{JHEP} {\bfseries 08}
  (2022) 237} [\href{https://arxiv.org/abs/2204.05331}{{\ttfamily
  2204.05331}}].

\bibitem{Marchesano:2021gyv}
F.~Marchesano, D.~Prieto and M.~Wiesner, \emph{{F-theory flux vacua at large
  complex structure}},
  \href{https://doi.org/10.1007/JHEP08(2021)077}{\emph{JHEP} {\bfseries 08}
  (2021) 077} [\href{https://arxiv.org/abs/2105.09326}{{\ttfamily
  2105.09326}}].

\bibitem{Lust:2021xds}
S.~L\"ust, \emph{{Large complex structure flux vacua of IIB and the Tadpole
  Conjecture}},  \href{https://arxiv.org/abs/2109.05033}{{\ttfamily
  2109.05033}}.

\bibitem{Grimm:2021ckh}
T.W.~Grimm, E.~Plauschinn and D.~van~de Heisteeg, \emph{{Moduli stabilization
  in asymptotic flux compactifications}},
  \href{https://doi.org/10.1007/JHEP03(2022)117}{\emph{JHEP} {\bfseries 03}
  (2022) 117} [\href{https://arxiv.org/abs/2110.05511}{{\ttfamily
  2110.05511}}].

\bibitem{Braun:2020jrx}
A.P.~Braun and R.~Valandro, \emph{{$G_{4}$ flux, algebraic cycles and complex
  structure moduli stabilization}},
  \href{https://doi.org/10.1007/JHEP01(2021)207}{\emph{JHEP} {\bfseries 01}
  (2021) 207} [\href{https://arxiv.org/abs/2009.11873}{{\ttfamily
  2009.11873}}].

\bibitem{Becker:2006ks}
K.~Becker, M.~Becker, C.~Vafa and J.~Walcher, \emph{{Moduli Stabilization in
  Non-Geometric Backgrounds}},
  \href{https://doi.org/10.1016/j.nuclphysb.2007.01.034}{\emph{Nucl. Phys. B}
  {\bfseries 770} (2007) 1}
  [\href{https://arxiv.org/abs/hep-th/0611001}{{\ttfamily hep-th/0611001}}].

\bibitem{Becker:2007dn}
K.~Becker, M.~Becker and J.~Walcher, \emph{{Runaway in the Landscape}},
  \href{https://doi.org/10.1103/PhysRevD.76.106002}{\emph{Phys. Rev. D}
  {\bfseries 76} (2007) 106002}
  [\href{https://arxiv.org/abs/0706.0514}{{\ttfamily 0706.0514}}].

\bibitem{Ishiguro:2021csu}
K.~Ishiguro and H.~Otsuka, \emph{{Sharpening the boundaries between flux
  landscape and swampland by tadpole charge}},
  \href{https://doi.org/10.1007/JHEP12(2021)017}{\emph{JHEP} {\bfseries 12}
  (2021) 017} [\href{https://arxiv.org/abs/2104.15030}{{\ttfamily
  2104.15030}}].

\bibitem{Bardzell:2022jfh}
J.~Bardzell, E.~Gonzalo, M.~Rajaguru, D.~Smith and T.~Wrase, \emph{{Type IIB
  flux compactifications with h$^{1,1}$ = 0}},
  \href{https://doi.org/10.1007/JHEP06(2022)166}{\emph{JHEP} {\bfseries 06}
  (2022) 166} [\href{https://arxiv.org/abs/2203.15818}{{\ttfamily
  2203.15818}}].

\bibitem{Acharya:2006zw}
B.S.~Acharya and M.R.~Douglas, \emph{{A Finite landscape?}},
  \href{https://arxiv.org/abs/hep-th/0606212}{{\ttfamily hep-th/0606212}}.

\bibitem{Vafa:2005ui}
C.~Vafa, \emph{{The String landscape and the swampland}},
  \href{https://arxiv.org/abs/hep-th/0509212}{{\ttfamily hep-th/0509212}}.

\bibitem{DeWolfe:2005uu}
O.~DeWolfe, A.~Giryavets, S.~Kachru and W.~Taylor, \emph{{Type IIA moduli
  stabilization}},
  \href{https://doi.org/10.1088/1126-6708/2005/07/066}{\emph{JHEP} {\bfseries
  07} (2005) 066} [\href{https://arxiv.org/abs/hep-th/0505160}{{\ttfamily
  hep-th/0505160}}].

\bibitem{Camara:2005dc}
P.G.~Camara, A.~Font and L.E.~Ibanez, \emph{{Fluxes, moduli fixing and
  MSSM-like vacua in a simple IIA orientifold}},
  \href{https://doi.org/10.1088/1126-6708/2005/09/013}{\emph{JHEP} {\bfseries
  09} (2005) 013} [\href{https://arxiv.org/abs/hep-th/0506066}{{\ttfamily
  hep-th/0506066}}].

\bibitem{Hori_2008}
K.~Hori and J.~Walcher, \emph{D-brane categories for
  orientifolds{\textemdash}the landau-ginzburg case},
  \href{https://doi.org/10.1088/1126-6708/2008/04/030}{\emph{Journal of High
  Energy Physics} {\bfseries 2008} (2008) 030}.

\bibitem{Dasgupta:1999ss}
K.~Dasgupta, G.~Rajesh and S.~Sethi, \emph{{M theory, orientifolds and G -
  flux}}, \href{https://doi.org/10.1088/1126-6708/1999/08/023}{\emph{JHEP}
  {\bfseries 08} (1999) 023}
  [\href{https://arxiv.org/abs/hep-th/9908088}{{\ttfamily hep-th/9908088}}].

\bibitem{Gukov:1999ya}
S.~Gukov, C.~Vafa and E.~Witten, \emph{{CFT's from Calabi-Yau four folds}},
  \href{https://doi.org/10.1016/S0550-3213(00)00373-4}{\emph{Nucl. Phys. B}
  {\bfseries 584} (2000) 69}
  [\href{https://arxiv.org/abs/hep-th/9906070}{{\ttfamily hep-th/9906070}}].

\bibitem{Kim:2022jvv}
M.~Kim, \emph{{D-instanton superpotential in string theory}},
  \href{https://doi.org/10.1007/JHEP03(2022)054}{\emph{JHEP} {\bfseries 03}
  (2022) 054} [\href{https://arxiv.org/abs/2201.04634}{{\ttfamily
  2201.04634}}].

\bibitem{Marino:1999af}
M.~Marino, R.~Minasian, G.W.~Moore and A.~Strominger, \emph{{Nonlinear
  instantons from supersymmetric p-branes}},
  \href{https://doi.org/10.1088/1126-6708/2000/01/005}{\emph{JHEP} {\bfseries
  01} (2000) 005} [\href{https://arxiv.org/abs/hep-th/9911206}{{\ttfamily
  hep-th/9911206}}].

\bibitem{Grimm:2004uq}
T.W.~Grimm and J.~Louis, \emph{{The Effective action of N = 1 Calabi-Yau
  orientifolds}},
  \href{https://doi.org/10.1016/j.nuclphysb.2004.08.005}{\emph{Nucl. Phys. B}
  {\bfseries 699} (2004) 387}
  [\href{https://arxiv.org/abs/hep-th/0403067}{{\ttfamily hep-th/0403067}}].

\bibitem{Grimm:2004ua}
T.W.~Grimm and J.~Louis, \emph{{The Effective action of type IIA Calabi-Yau
  orientifolds}},
  \href{https://doi.org/10.1016/j.nuclphysb.2005.04.007}{\emph{Nucl. Phys. B}
  {\bfseries 718} (2005) 153}
  [\href{https://arxiv.org/abs/hep-th/0412277}{{\ttfamily hep-th/0412277}}].

\bibitem{Candelas:1990pi}
P.~Candelas and X.~de~la Ossa, \emph{{Moduli Space of {Calabi-Yau} Manifolds}},
  \href{https://doi.org/10.1016/0550-3213(91)90122-E}{\emph{Nucl. Phys. B}
  {\bfseries 355} (1991) 455}.

\bibitem{Denef_2004}
F.~Denef and M.R.~Douglas, \emph{Distributions of flux vacua},
  \href{https://doi.org/10.1088/1126-6708/2004/05/072}{\emph{Journal of High
  Energy Physics} {\bfseries 2004} (2004) 072}.

\bibitem{Denef_2007}
F.~Denef and M.R.~Douglas, \emph{Computational complexity of the landscape:
  Part i}, \href{https://doi.org/10.1016/j.aop.2006.07.013}{\emph{Annals of
  Physics} {\bfseries 322} (2007) 1096}.

\bibitem{Becker:2024ijy}
K.~Becker, M.~Rajaguru, A.~Sengupta, J.~Walcher and T.~Wrase,
  \emph{{Stabilizing massless fields with fluxes in Landau-Ginzburg models}},
  \href{https://doi.org/10.1007/JHEP08(2024)069}{\emph{JHEP} {\bfseries 08}
  (2024) 069} [\href{https://arxiv.org/abs/2406.03435}{{\ttfamily
  2406.03435}}].

\bibitem{Becker:2024ayh}
K.~Becker, N.~Brady, M.~Gra\~na, M.~Morros, A.~Sengupta and Q.~You,
  \emph{{Tadpole conjecture in non-geometric backgrounds}},
  \href{https://doi.org/10.1007/JHEP10(2024)021}{\emph{JHEP} {\bfseries 10}
  (2024) 021} [\href{https://arxiv.org/abs/2407.16758}{{\ttfamily
  2407.16758}}].

\bibitem{Rajaguru:2024emw}
M.~Rajaguru, A.~Sengupta and T.~Wrase, \emph{{Fully stabilized Minkowski vacua
  in the 2$^{6}$ Landau-Ginzburg model}},
  \href{https://doi.org/10.1007/JHEP10(2024)095}{\emph{JHEP} {\bfseries 10}
  (2024) 095} [\href{https://arxiv.org/abs/2407.16756}{{\ttfamily
  2407.16756}}].

\bibitem{Andriot:2022yyj}
D.~Andriot, L.~Horer and P.~Marconnet, \emph{{Exploring the landscape of
  (anti-) de Sitter and Minkowski solutions: group manifolds, stability and
  scale separation}},
  \href{https://doi.org/10.1007/JHEP08(2022)109}{\emph{JHEP} {\bfseries 08}
  (2022) 109} [\href{https://arxiv.org/abs/2204.05327}{{\ttfamily
  2204.05327}}].

\bibitem{Grimm:2021vpn}
T.W.~Grimm, \emph{{Taming the Landscape of Effective Theories}},
  \href{https://arxiv.org/abs/2112.08383}{{\ttfamily 2112.08383}}.

\bibitem{Bakker:2021uqw}
B.~Bakker, T.W.~Grimm, C.~Schnell and J.~Tsimerman, \emph{{Finiteness for
  self-dual classes in integral variations of Hodge structure}},
  \href{https://arxiv.org/abs/2112.06995}{{\ttfamily 2112.06995}}.

\bibitem{Hamada:2021yxy}
Y.~Hamada, M.~Montero, C.~Vafa and I.~Valenzuela, \emph{{Finiteness and the
  swampland}}, \href{https://doi.org/10.1088/1751-8121/ac6404}{\emph{J. Phys.
  A} {\bfseries 55} (2022) 224005}
  [\href{https://arxiv.org/abs/2111.00015}{{\ttfamily 2111.00015}}].

\bibitem{Breitenlohner:1982bm}
P.~Breitenlohner and D.Z.~Freedman, \emph{{Positive Energy in anti-De Sitter
  Backgrounds and Gauged Extended Supergravity}},
  \href{https://doi.org/10.1016/0370-2693(82)90643-8}{\emph{Phys. Lett. B}
  {\bfseries 115} (1982) 197}.

\bibitem{Hanany:1997sa}
A.~Hanany and A.~Zaffaroni, \emph{{Chiral symmetry from type IIA branes}},
  \href{https://doi.org/10.1016/S0550-3213(97)00595-6}{\emph{Nucl. Phys. B}
  {\bfseries 509} (1998) 145}
  [\href{https://arxiv.org/abs/hep-th/9706047}{{\ttfamily hep-th/9706047}}].

\bibitem{Brunner:1997gf}
I.~Brunner and A.~Karch, \emph{{Branes at orbifolds versus Hanany Witten in
  six-dimensions}},
  \href{https://doi.org/10.1088/1126-6708/1998/03/003}{\emph{JHEP} {\bfseries
  03} (1998) 003} [\href{https://arxiv.org/abs/hep-th/9712143}{{\ttfamily
  hep-th/9712143}}].

\bibitem{Ahn:1998pb}
C.~Ahn, K.~Oh and R.~Tatar, \emph{{Orbifolds of AdS(7) x S**4 and
  six-dimensional (0,1) SCFT}},
  \href{https://doi.org/10.1016/S0370-2693(98)01276-3}{\emph{Phys. Lett. B}
  {\bfseries 442} (1998) 109}
  [\href{https://arxiv.org/abs/hep-th/9804093}{{\ttfamily hep-th/9804093}}].

\bibitem{Hanany:1997gh}
A.~Hanany and A.~Zaffaroni, \emph{{Branes and six-dimensional supersymmetric
  theories}}, \href{https://doi.org/10.1016/S0550-3213(98)00355-1}{\emph{Nucl.
  Phys. B} {\bfseries 529} (1998) 180}
  [\href{https://arxiv.org/abs/hep-th/9712145}{{\ttfamily hep-th/9712145}}].

\bibitem{Ferrara:1998vf}
S.~Ferrara, A.~Kehagias, H.~Partouche and A.~Zaffaroni, \emph{{Membranes and
  five-branes with lower supersymmetry and their AdS supergravity duals}},
  \href{https://doi.org/10.1016/S0370-2693(98)00558-9}{\emph{Phys. Lett. B}
  {\bfseries 431} (1998) 42}
  [\href{https://arxiv.org/abs/hep-th/9803109}{{\ttfamily hep-th/9803109}}].

\bibitem{Lust:2019zwm}
D.~L\"ust, E.~Palti and C.~Vafa, \emph{{AdS and the Swampland}},
  \href{https://doi.org/10.1016/j.physletb.2019.134867}{\emph{Phys. Lett. B}
  {\bfseries 797} (2019) 134867}
  [\href{https://arxiv.org/abs/1906.05225}{{\ttfamily 1906.05225}}].

\bibitem{Veneziano:2001ah}
G.~Veneziano, \emph{{Large N bounds on, and compositeness limit of, gauge and
  gravitational interactions}},
  \href{https://doi.org/10.1088/1126-6708/2002/06/051}{\emph{JHEP} {\bfseries
  06} (2002) 051} [\href{https://arxiv.org/abs/hep-th/0110129}{{\ttfamily
  hep-th/0110129}}].

\bibitem{Arkani-Hamed:2005zuc}
N.~Arkani-Hamed, S.~Dimopoulos and S.~Kachru, \emph{{Predictive landscapes and
  new physics at a TeV}},
  \href{https://arxiv.org/abs/hep-th/0501082}{{\ttfamily hep-th/0501082}}.

\bibitem{Dvali:2007hz}
G.~Dvali, \emph{{Black Holes and Large N Species Solution to the Hierarchy
  Problem}}, \href{https://doi.org/10.1002/prop.201000009}{\emph{Fortsch.
  Phys.} {\bfseries 58} (2010) 528}
  [\href{https://arxiv.org/abs/0706.2050}{{\ttfamily 0706.2050}}].

\bibitem{Dvali:2007wp}
G.~Dvali and M.~Redi, \emph{{Black Hole Bound on the Number of Species and
  Quantum Gravity at LHC}},
  \href{https://doi.org/10.1103/PhysRevD.77.045027}{\emph{Phys. Rev. D}
  {\bfseries 77} (2008) 045027}
  [\href{https://arxiv.org/abs/0710.4344}{{\ttfamily 0710.4344}}].

\bibitem{Berglund_1994}
P.~Berglund, P.~Candelas, X.~de~la Ossa, A.~Font, T.~Hübsch,
  D.~Jan{\v{c}}i{\'{c}} et~al., \emph{Periods for {C}alabi-{Y}au and
  {L}andau-{G}inzburg vacua},
  \href{https://doi.org/10.1016/0550-3213(94)90047-7}{\emph{Nuclear Physics B}
  {\bfseries 419} (1994) 352}.

\end{thebibliography}\endgroup

\end{document}